\documentclass[10pt,journal,compsoc]{IEEEtran}

\usepackage{ifpdf}
\usepackage{url}
\usepackage[nocompress]{cite}
\usepackage[caption=false,font=footnotesize,labelfont=sf,textfont=sf]{subfig}
\usepackage{hyperref}
\usepackage{ulem}
\ifCLASSINFOpdf
	\usepackage[pdftex]{graphicx}
	\DeclareGraphicsExtensions{.pdf}
\else
	\usepackage[dvips]{graphicx}
	\DeclareGraphicsExtensions{.eps}
\fi

\graphicspath{{./figures/}}

\begin{document}

\title{syGlass: Interactive Exploration of Multidimensional Images Using Virtual Reality Head-mounted Displays}

\author{Stanislav~Pidhorskyi$^1$,~\IEEEmembership{Student Member,~IEEE} ,
        Michael~Morehead$^1$, 
        Quinn~Jones$^1$,~\IEEEmembership{Student Member,~IEEE},
        George~Spirou$^2$, 
        and~Gianfranco~Doretto$^1$,~\IEEEmembership{Member,~IEEE}}

%\markboth%
%{Pidhorskyi \MakeLowercase{\textit{et al.}}: syGlass: Interactive Exploration of Multidimensional Images Using Virtual Reality Head-mounted Displays}

\IEEEtitleabstractindextext{%
\begin{abstract}
The quest for deeper understanding of biological systems has driven the acquisition of increasingly larger multidimensional image datasets. Inspecting and manipulating data of this complexity is very challenging in traditional visualization systems. We developed syGlass, a software package capable of visualizing large scale volumetric data with inexpensive virtual reality head-mounted display technology. This allows leveraging stereoscopic vision to significantly improve perception of complex 3D structures, and provides immersive interaction with data directly in 3D. We accomplished this by developing highly optimized data flow and volume rendering pipelines, tested on datasets up to 16TB in size, as well as tools available in a virtual reality GUI to support advanced data exploration, annotation, and cataloguing.
\end{abstract}

}

\maketitle

\footnotetext[1]{Lane Department of Computer Science and Electrical Engineering of West Virginia.University}
\footnotetext[2]{Rockefeller Neuroscience Institute of West Virginia University}

\IEEEraisesectionheading{\section{Introduction}\label{sec:introduction}}

\IEEEPARstart{I}{creasingly}, medical diagnoses and biology research
breakthroughs rely on multidimensional imaging technologies. In medicine, well-known
examples include Magnetic Resonance Imaging (MRI) and Computerized Axial
Tomography (CAT).  In biology, specifically structural biology and neuroscience,
different forms of microscopy imaging, including Serial Block-Face Scanning
Electron Microscopy (SBFSEM), 2-photon Microscopy, Lattice Light-Sheet
Microscopy (LLSM) and others, are routinely used to acquire datasets approaching
tens of TB in size. Indeed, SBFSEM has been applied on increasingly large volumes reconstructing 3D models of circuits, cells and organelles. LLSM, instead, allows collecting three dimensional images at sub-second intervals on living tissue~\cite{Chen2014}, thus creating 3D movies featuring cells moving and growing.
\cite{lengyel2010game}
To effectively leverage multidimensional imaging technologies and potentially
reach a deeper understanding of biological systems, researchers need the right
tools for exploring, conceptualizing, and annotating data that is challenging
because it is intrinsically complex and vast in size. Its volumetric nature makes it hard to analyze and work with through traditional visualization and input systems. It is difficult to discern tangled 3D structures by rendering an image volume slice by slice, while projecting the volume in 2D leads to information loss. 

In this work, we present \textit{syGlass} (\url{www.syglass.io}), a
first-of-its-kind software package for the interactive exploration of large-scale
multidimensional images, enabling the user to exploit the power of
binocular vision, since it leverages modern virtual reality (VR) hardware
technology. Until very recently, immersive VR was experienced through the use of
costly and static CAVE-like installations~\cite{moreheadJBHSDEDS143dui}, but
syGlass has been developed for current head-mounted display (HMD) technology,
which is inexpensive and portable, yet with a wide field of view, accurate low latency tracking, and high-resolution displays that guarantee comfortable depth perception in an immersive experience. Moreover, HMD technology includes input devices with 6 degrees of freedom that far surpass the ability to interact with image volumes provided by regular computer mice. For more on bioimaging software tools, see the sidebar.

Using HMD's implies using only a stand-alone workstation with limited computing
power for visualizing large scale image volumes at high frame rates to avoid VR
sickness effects~\cite{sickness}. To address that challenge, in syGlass we have
developed highly optimized data flow and volume rendering pipelines. In
addition, syGlass provides a custom GUI with a variety of tools for interacting
with the data in VR, for enhancing the visualization of image volumes according
to the user's needs, and for performing several types of complex data annotation and cataloging, directly in VR.

%%% Local Variables:
%%% mode: latex
%%% TeX-master: "syglasspaper-main"
%%% End:

%\section{Related work}

\section{Related Work in Bioimaging Software}

There are several bioimaging software packages that provide multidimensional image visualization and analysis. Some prominent examples from the open-source and academic community include Vaa3D~\cite{Peng2014}, NeuroBlocks~\cite{AiAwami2016}, and Catmaid~\cite{Saalfeld2009}. In the proprietary category, popular tools are Imaris (\url{www.bitplane.com}) and Amira (\url{www.fei.com}). These packages provide specialized analysis tools, often with orthogonal strengths, and generic visualization capabilities. However, they are limited because data is visualized either by allowing the user to view and scroll through individual parallel slices of the volume or as a 2D projection. 

The main difference with previous packages is that syGlass enables the user to ``see'' and operate in VR, since immersive mediums have been shown to improve users' ability to conceptualize 3D data~\cite{Laha2012}. While Amira offers the option for working in VR, according to their promotional material, they are compatible with older CAVE systems, which take up entire rooms and are much more expensive than HMD's, which syGlass is designed to use. 

However, relying on HMD's presents new challenges,
which the aforementioned packages do not have to consider. The most predominant
being that HMD's have a high screen resolution and require high rendering frame
rates simultaneously in both eyes to avoid user discomfort, like motion sickness. Coupling
that with the limited computing power of a stand-alone workstation driving the HMD, as well as the need to handle TB's of data, poses technical problems difficult to address, even for seasoned brands in this industry.

\section{syGlass Overview}

syGlass is a complex system that at high level includes the following components:
\begin{itemize}
  \item Native Visualization and Annotation Application (NVAA)
  \item Python-based Server Application
  \item Electron-based Manager Application (syBook)
  \item Local and Remote  Annotation Databases
  \item Local Volumetric Data Storage
\end{itemize}

\begin{figure}[t!]
\centering
\includegraphics[width=3.5in]{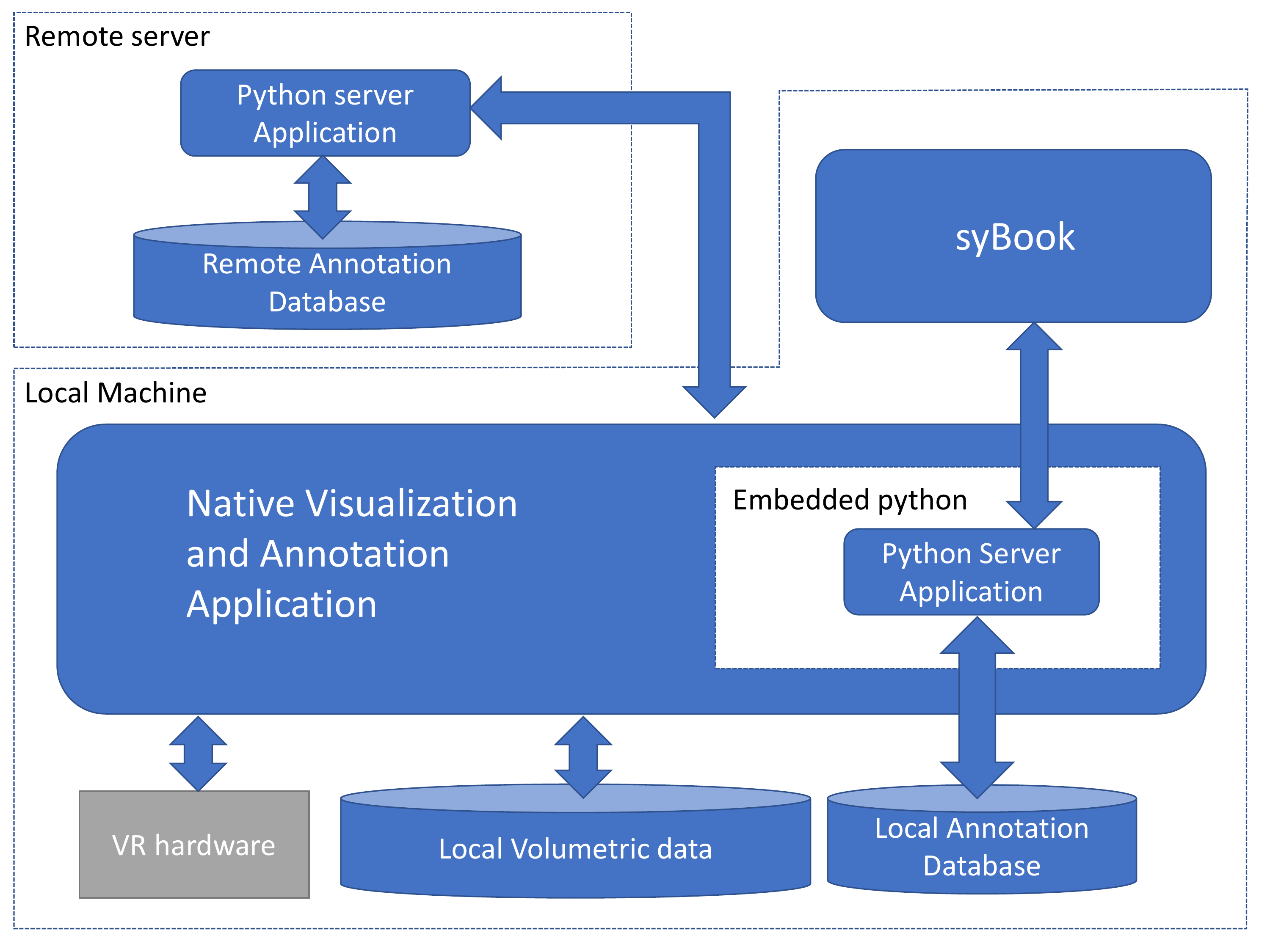}
\caption{\textbf{syGlass overview.} The system consists of a native application for image data visualization and annotation, and a server python-based application for storing annotation data. The latter can run in standalone mode on a remote server, and it also runs embedded in the native application.}
\label{fig:top_level}
\end{figure}

The NVAA loads and visualizes volumetric image data with direct volume rendering
(DVR), loads and visualizes mesh and annotation data, and provides annotation
tools in VR. Volumetric data is stored locally in custom binary containers,
which are optimized for fast data transfer with the GPU. Mesh and annotation
data are stored in the Annotation Database, operated by the Python-based Server
Application. The database can be operated locally or remotely, for performing
individual or collaborative work. The local instance is operated by the same
Python-based Server Application, which runs in the NVAA with an embedded Python environment to allow for a unified interface. Raw data (either mesh or volumetric type of data) is ingested by syGlass in the form of \textit{project} instances. Besides raw data, projects contain annotation data. Project management is performed by \textit{syBook}, an Electron-based Manager Application. Electron (\url{https://electron.atom.io}), is a GUI framework for running JavaScript code. The interaction between all of these components is shown in Figure~\ref{fig:top_level}.

\begin{figure*}[th!]
\centering
\includegraphics[width=7.0in]{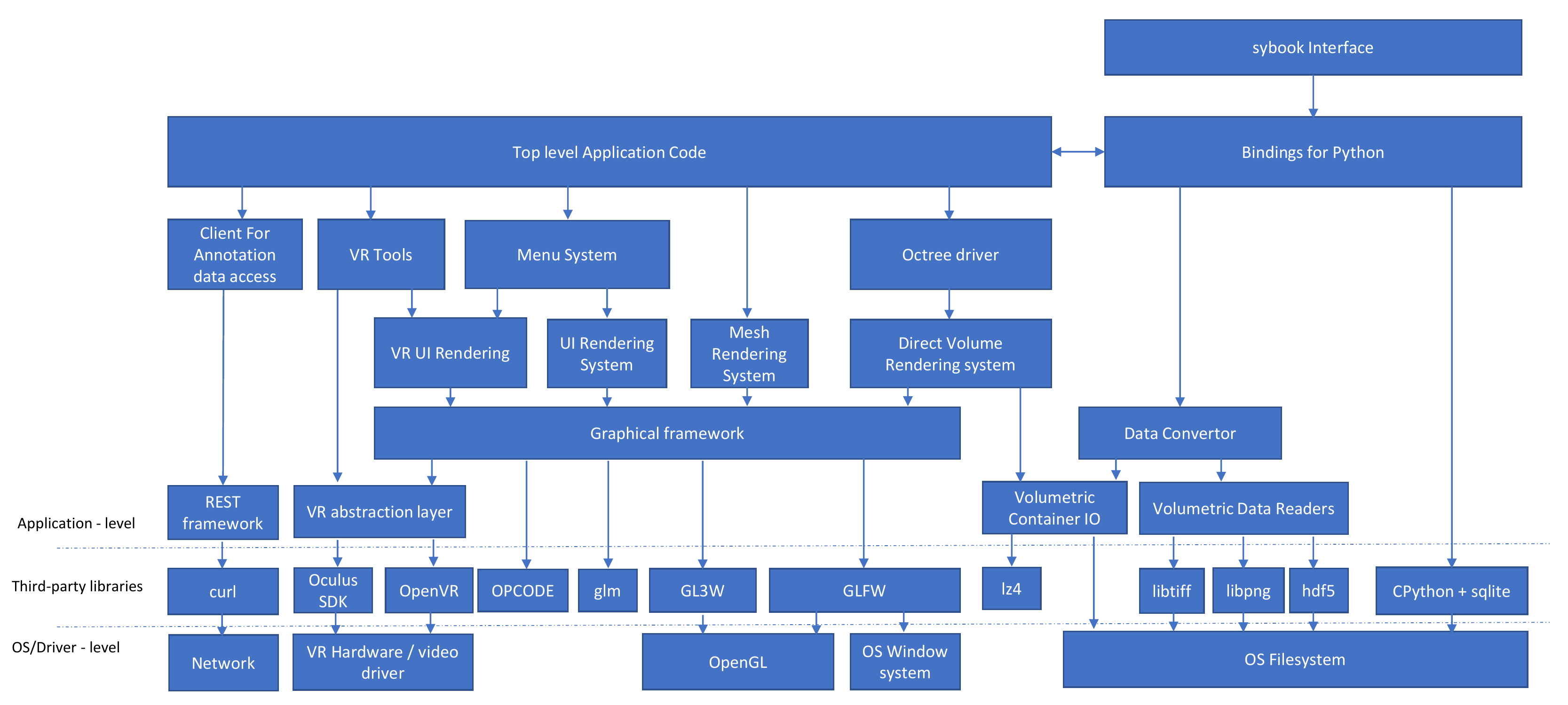}
\caption{\textbf{Native Visualization and Annotation Application overview.} The hierarchy components organized by OS/Driver, Third-party library, and Application levels, as well as by their function within each level.}
\label{fig:structure}
\end{figure*}

Figure~\ref{fig:structure} shows the structure of NVAA. It is a cross-platform application that can run on Microsoft Windows 7, 8 and 10, GNU-Linux (tested on Ubuntu 14.04 LTS and 16.04 LTS), and OS X (tested on 10.9 and 10.10). However, current active support is focused on Microsoft Windows due to the lack of third-party support of the other platforms by HMD manufacturers. The code base is C++14 compliant and can be compiled by MSVC 14.0, GCC 5, Clang 3.4, or later versions of any of the aforementioned. Native makefiles and workspace for each platform are generated from independent configuration files using CMake, while building of resources is performed by the platform-independent build system SCons. Build automation for production is performed with Jenkins.

\subsection{VR Abstraction Layer}

\begin{figure}[th!]
\centering
\includegraphics[width=3.5in]{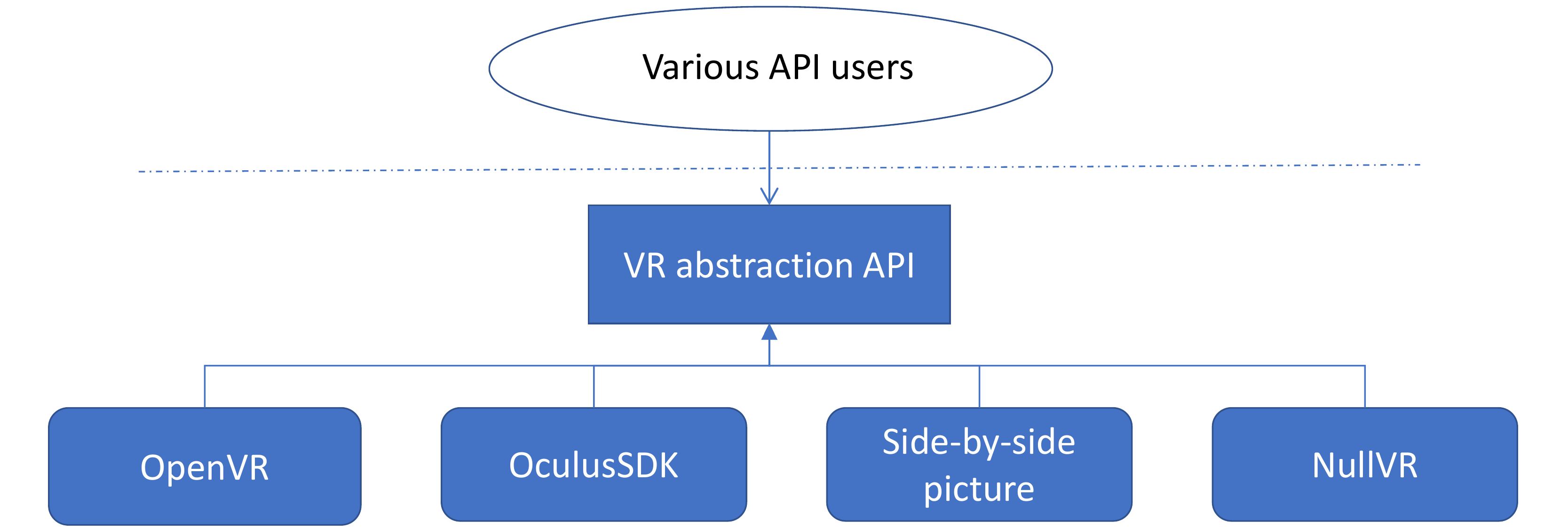}
\caption{\textbf{VR Abstraction Layer.} The VR Abstraction Layer provides a
  unified API for the NVAA. Abstraction implementations are provided for Oculus SDK, OpenVR, on-monitor side-by-side picture, and NullVR as a degenerate case.}
\label{fig:vr_api}
\end{figure}

We provide support for two VR HMD systems: the Oculus Rift and the HTC Vive. The
Vive is supported through OpenVR, which is a runtime library aimed at providing access to HMD hardware from multiple vendors. OpenVR is a component under SteamVR and requires the installation of Steam, a video game digital rights management service. We also provide native support for the Rift through the Oculus SDK. Thus, installing SteamVR is not necessary if the user intends to use syGlass with a Rift. To ensure seamless support of different VR API's, we implemented a layer of abstraction for VR which provides a unified API for the rest of the application. See Figure~\ref{fig:vr_api}. There are several implementations of the VR abstraction API, including for the Oculus SDK, for OpenVR, for on-monitor side-by-side picture, and for NullVR, which handles degenerate cases like the absence of HMD hardware.

\subsection{Graphical API}

OpenGL is the only graphics API that we utilize, due to its cross-platform nature and widespread use. The required version is 4.3 (Release date: July 22, 2013), because it includes compute shaders. The application window and OpenGL context creation are handled by the GLFW library, while loading of the latest OpenGL core profile is done by GL3W.

All visualization tasks exploit the GPU (through fragment and compute shaders). However, we neither depend on CUDA, nor require a CUDA capable device. Volumetric rendering is performed in multiple stages to offscreen buffers using ray-casting in fragment or compute shaders. Geometric shaders are used to render data annotation primitives via imposters. In this way, the complex geometry for annotations are generated on GPU dynamically out of a smaller amount of input primitives, which can be updated very quickly when an edit operation is performed.

\subsection{Raw Data}

The raw input data can be either volumetric image data, or mesh data. Raw data is imported when a new project is created. Mesh data can be imported from obj files, and is stored in the Annotation Database. 
Volumetric data is stored locally and is imported into a custom container. Input volumetric formats can be various TIFF files, Imaris files, and sequences of png or jpeg files. The container file format is optimized for visualization purposes.

\subsection{Interaction with Raw and Annotation Data}

In VR, users interact with syGlass through controllers provided with the HMD
hardware, or to be bought separately, in the case of the Oculus Touch. Controllers provide 6 degrees of freedom for the input of position and orientation plus several buttons, triggers, joysticks or track-pads. Six degrees of freedom enable superior input for many operations that are inefficient with a regular computer mouse that has only 2 degrees. 

syGlass provides several "tools" represented in VR as physical tools operated by
the controllers, with which the user interacts with raw and annotation data.
User experience with the tools was optimized  to make it as natural as possible, or as familiar as possible. For instance, users can use a ``pinch-to-zoom'' like action to scale the data volume, borrowing from its now common use in smartphones interfaces.

Several tools provide interaction with raw data. The geometric transformation tool permits translation, rotation, and scaling. With one controller the user can translate and rotate a volume, and with two controllers the user can scale the volume in a pinch-to-zoom fashion. The cut-plane tool allows rendering a volume without a portion that has been ``cut-off''. The ROI tool draws a rectangular cuboidal region that is renderend or that should be used to query the Annotation Database for information.

Another set of tools operates on annotation data. There are tools for placing
markers, for counting objects, for tracking features in 3D movies, for measuring
distances, and for taking 2D images and recording 2D virtual movies from a
predefined vantage point or trajectory. In addition, there is also a tool for
drawing 3D graphs, designed for identifying the skeletal structure of biological processes. Annotation data can be exported, and is searchable through syBook.

%%% Local Variables:
%%% mode: latex
%%% TeX-master: "syglasspaper-main"
%%% End:

\section{Data Flow}

Raw data ingestion begins by generating a project using syBook. The project manager allows the creation of either mesh or volumetric projects, containing mesh or volumetric data, respectively. The manager also allows specifying several properties: tags, a description, and the raw data files to import.  

\subsection{Volumetric Data Storage}\label{container}

Volumetric projects are stored locally in two files. The first is an immutable container with volumetric data, generated by importing raw data files at project creation time. The second is a zip archive with a LevelDB database, which allows very fast updates of annotation data and amortizes disk operations.

The container consists of one file storing compressed chunks of volumetric data
at multiple resolutions. During the conversion process, the file is accessed in
append mode. This prevents previously written chunks from becoming corrupted if
the conversion is interrupted. Chunks have headers allowing syGlass to resume
the conversion from where it was stopped. An index table accounts for the
offsets to all chunks. Chunks are compressed with LZ4. After decompression, they
can be efficiently uploaded to GPU as a 3D texture, because the voxel data they represent is already memory aligned.

% \begin{figure}[th!]
% \centering
% \includegraphics[height=0.25\textwidth]{figure4}
% \caption{\textbf{Data conversion process.}}
% \label{fig:data_conversion}
% \end{figure}

\subsection{Volumetric Data Cashing}

We implemented two memory caches to amortize disk operations and data transfer from CPU to GPU. One is located on CPU memory, and the other on GPU memory. The strategy of both caches is to maintain an amount of data chunks as large as possible while removing the least used chunks. Chunks have a timestamp attribute, which is updated each time they are used. Each cache maintains a min-heap priority queue of chunks organized by timestamps. If the cache grows up to the maximum allowed size and a new chunk is requested, then space is freed up by deleting chunks with the least priority, i.e. the oldest chunks.

\subsection{Mesh Data Storage}

Mesh projects are stored in the Annotation Database. This is possible since mesh data have a much smaller storage footprint than volumetric data. Mesh data is typically segmented and is rather a collection of meshes that need to be kept organized. We do so by storing mesh relations in the SQL database. In particular, a hierarchy of meshes is maintained so that the top level represents a volume of tissue, which comprises of a set of cell/organism structures at the next level, each of which contains mesh coordinates representing cell/organism substructures at the bottom level.

%%% Local Variables:
%%% mode: latex
%%% TeX-master: "syglasspaper-main"
%%% End:

\section{Volume Rendering}

Volume visualization can be distinguished between Indirect Volume Rendering (IVR) and Direct Volume Rendering (DVR). IVR corresponds to a rendering of preprocessed data, which typically means surfaces extracted from the volume, either manually or with automated segmentation tools. syGlass supports IVR through rendering mesh data. It does not perform segmentation and an external tool should be used for data preprocessing.

Direct Volume Rendering (DVR) is a technique that allows rendering volumetric
(voxel-based) data that have not been preprocessed for IVR. Depending on the
application, usually there are obvious reasons for preferring to visualize raw data in volumetric form using DVR, versus preprocessing it and visualizing mesh data with IVR. Within syGlass we have developed an advanced DVR engine that has been tested to work with volumetric datasets reaching sizes up to 16TB. 

\subsection{DVR Techniques}\label{dvr_techniques}

In DVR, volumetric data is considered to be a semi-transparent, light-emitting medium. Rendering techniques are based on physical laws for light emission, absorption, and scattering. In syGlass, these are combined with ray casting, which provides good quality and is inherently parallel, making it very efficient for execution on GPU. Data is stored and sampled using 3D textures, for which hardware three-linear interpolation is available.

%#There are several DVR techniques, such as:

%\begin{itemize}
%	\item Texture Mapping
%	\item Shear Warp
%	\item Splatting
%	\item Ray Casting
%\end{itemize}

%The most flexible and artifact-free is the Ray Casting technique. For the needs of VR, the perspective projection must be used which makes Texture Mapping and Shear Warp not efficient. Splatting in general introduces a lot of artifacts and may not be efficiently implemented for parallel execution on GPU.

In ray casting rendering happens on a per-pixel basis. For a pixel with intensity $I_p$, in position $p$ in a 2D image, a ray is cast through the volume hosting the medium $f(x)$ at voxel position $x$, which is sampled along the ray with some step size. The relation between pixel intensities and voxel values (i.e., the \textit{optical model} \cite{Max1995}) in syGlass does not consider scattering effects because it is very computationally expensive, thus unsuited for interactive immersive VR with HMD's. We use instead the emission-only model, given by
\[I_p = \int_{x_0}^{x_1}c(x)dx \; ,\]
where $c(x) = \tau[f(x)]$ is the intensity emitted by the voxel at position $x$, $x_0$ is the entry point and $x_1$ the exit point of the ray. The map $\tau[ \cdot ]$ is a so-called \textit{transfer function}, for which syGlass provides multiple options, aimed at generating different rendering of the volumetric data. Note that the contribution $c(x)$ at voxel $x$ to the final pixel value $I_p$ does not depend on other voxels between the viewer and the given voxel.
\begin{figure}[t!]
\begin{tabular}{cc}
\subfloat[]{\includegraphics[width=1.6in]{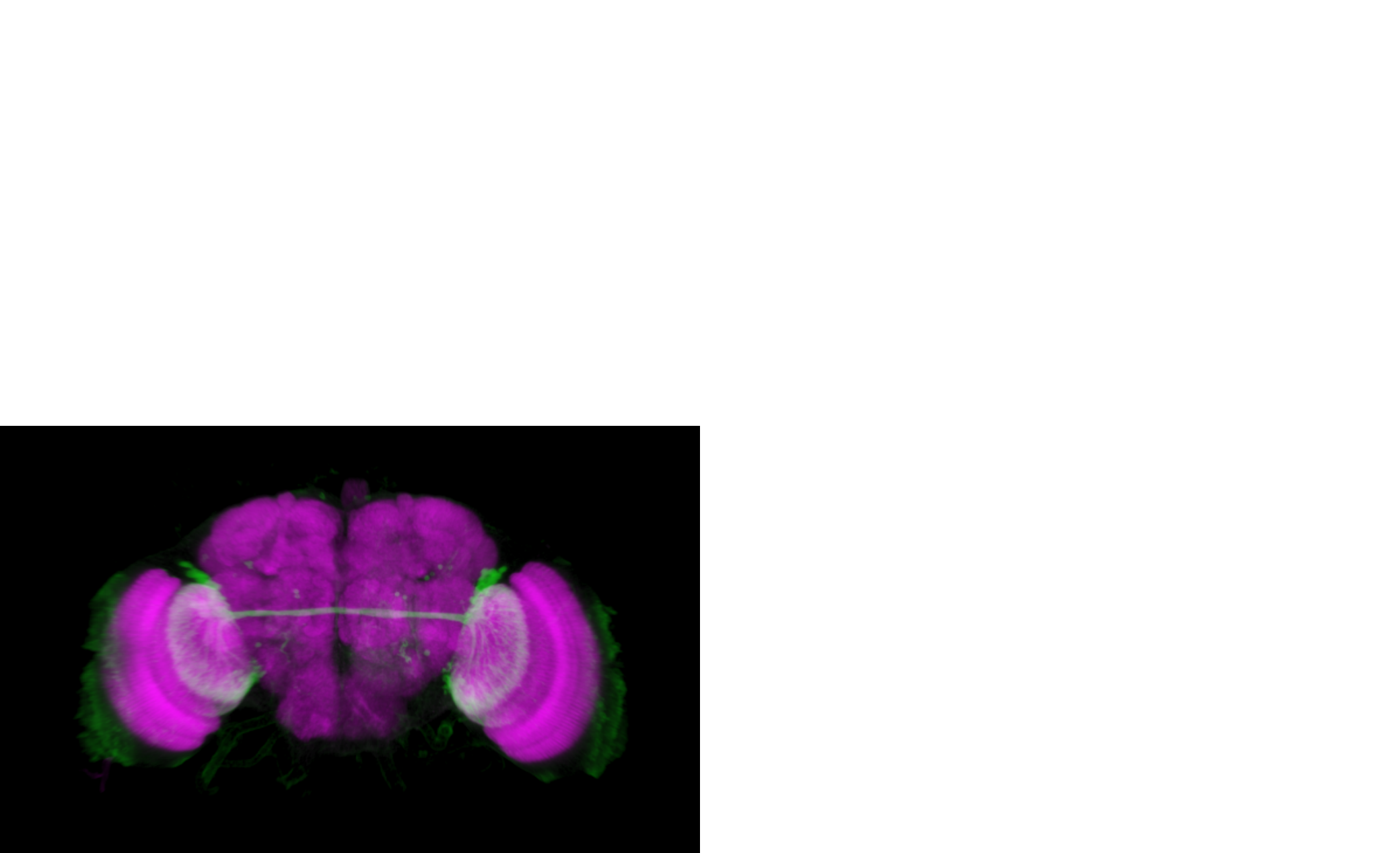}} & \subfloat[]{\includegraphics[width=1.6in]{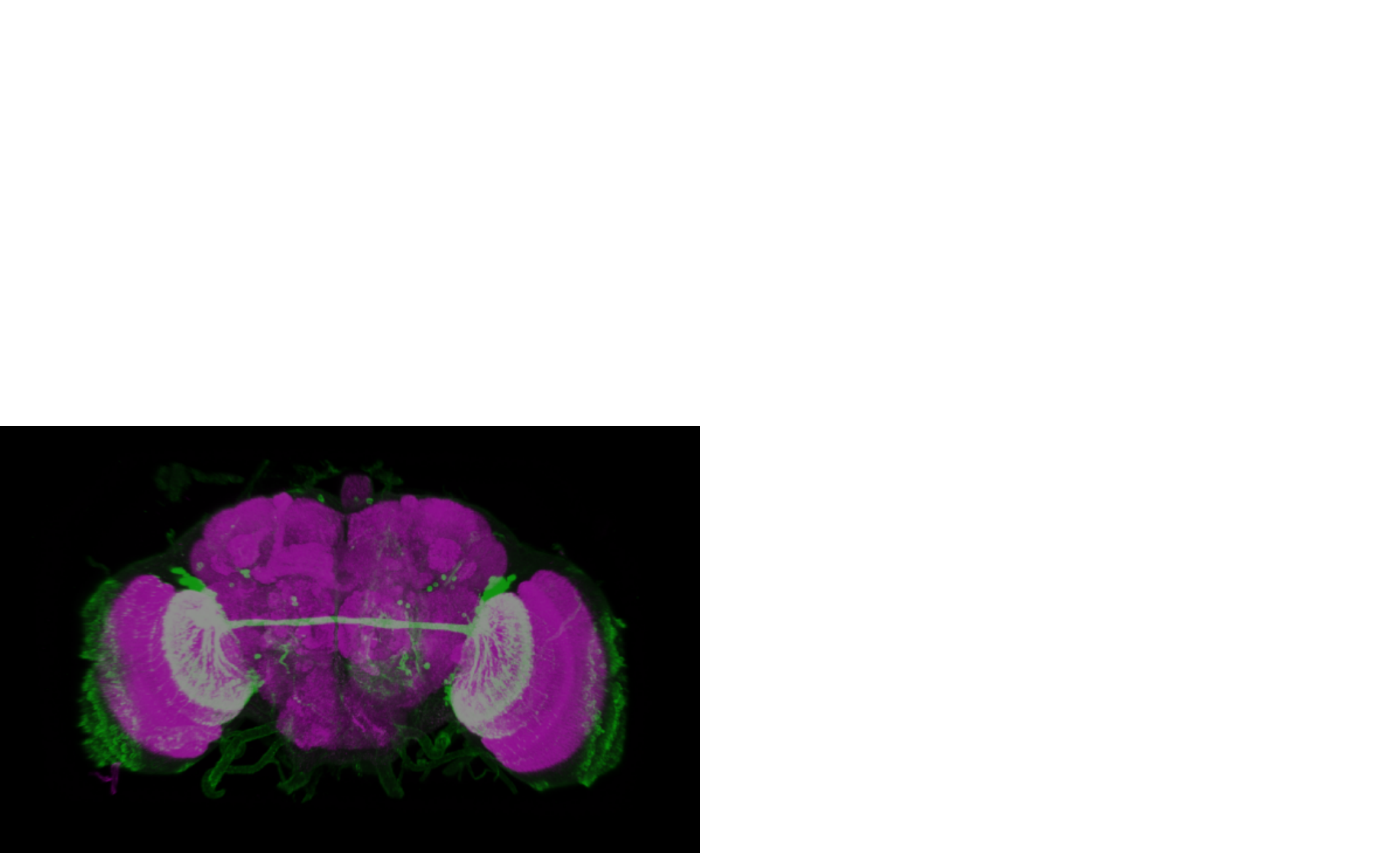}} \\
\subfloat[]{\includegraphics[width=1.6in]{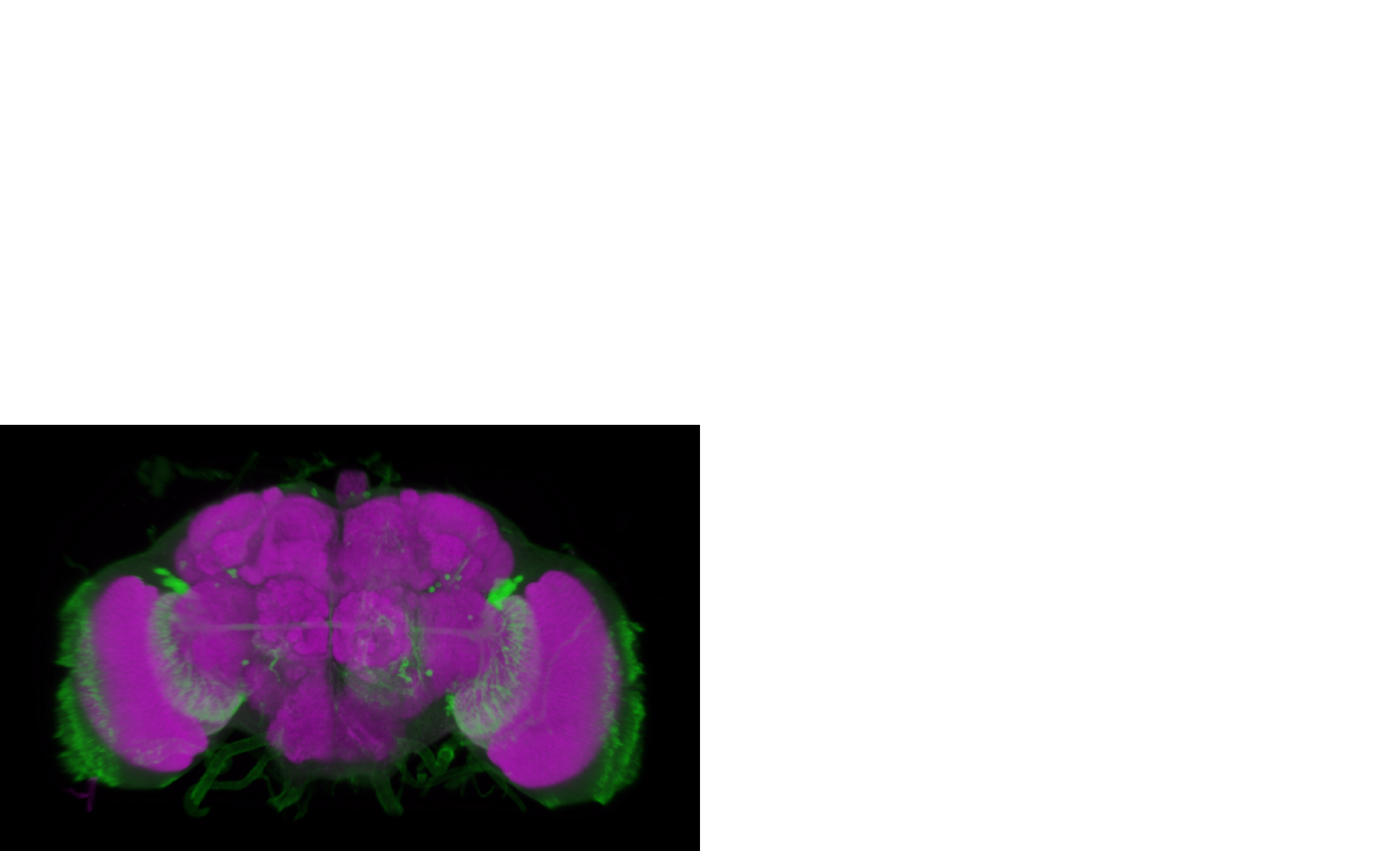}} & \subfloat[]{\includegraphics[width=1.6in]{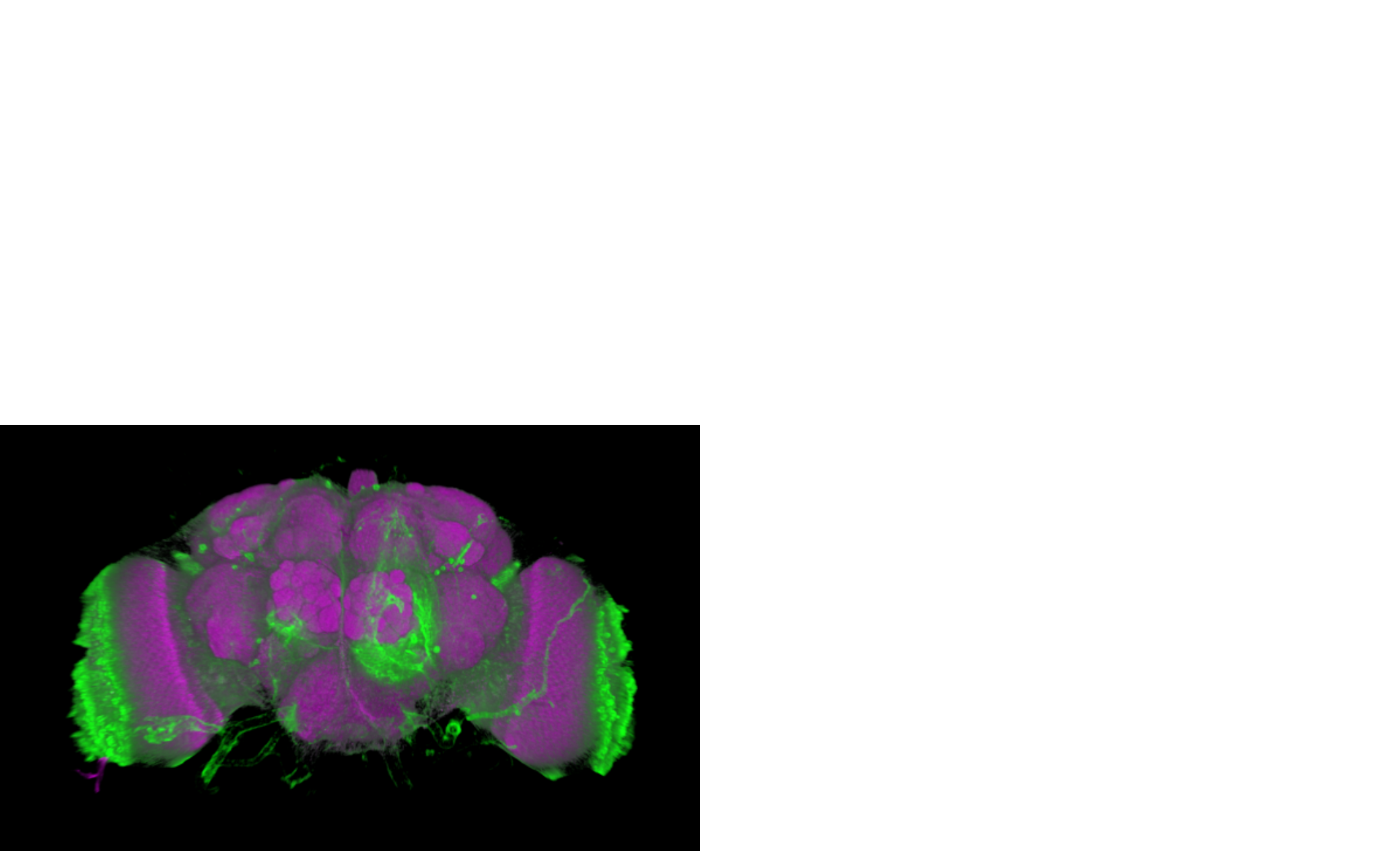}}
\end{tabular}
\caption{\textbf{Optical models comparison.} a) Emission-only model; b) Maximum intensity projection model; c) Emission-absorption model with low opacity; d) Emission-absorption model with high opacity. The volumetric data of the adult \textit{Drosophila} brain~\cite{Peng2014} has two channels: purple and green. Different models highlight different aspects of the 3D structure of the data.}
\label{fig:optical_models}
\end{figure}

While the emission-only model can be useful for sparse data, it is not revealing for dense data, when a lot of different objects contribute to the same pixel. In that case, the resulting image becomes a blend of many objects, making it hard to interpret. To better handle denser volumetric datasets, syGlass also has an emission-absorption model, given by
\[I_p = \int_{x_0}^{x_1}c(x)e^{-\alpha(x)}dx \; , \quad \alpha(x)= \int_{x_0}^{x}\rho(y)dy \; ,\]
where $\alpha$ is the accumulated opacity along the ray, and $\rho(y) = T[f(y)]$ is the opacity at voxel position $y$. The map $T[ \cdot ]$ is also a transfer function, for which syGlass provides multiple options. In this model, the contribution of a given voxel depends on the opacity of the voxels between the given voxel and the viewer. Note that when $e^{-\alpha(x)}$ becomes sufficiently small, the ray can be terminated, whereas in the emission-only model the integration has to proceed until $x_1$. Figure~\ref{fig:optical_models} shows a comparison between optical models.

\subsection{Octree Representation}

Volumetric data easily exceeds the on-board GPU memory, and to handle large-scale volumes we implemented a scalable DVR pipeline based on storing and rendering data organized in \textit{octrees}. An octree structure allows storing data as blocks in a multi-resolution pyramid. The goal then becomes to select blocks containing voxels currently traversed, at levels of resolution of the pyramid that match the pixel resolution of the rendered images. This strategy allows us to maintain a constant amount of voxels sampled per frame, significantly lowering the amount of memory and computational resources needed to render the frame. With this pipeline syGlass was proven to visualize datasets of size up to 16TB without creating user discomfort, but the maximum dataset size that can be visualized is virtually unbound, given that computationally, it is limited by the overhead of the octree traversal discussed below.

\begin{figure}[t!]
\centering
\subfloat[]{\includegraphics[height=2.0in]{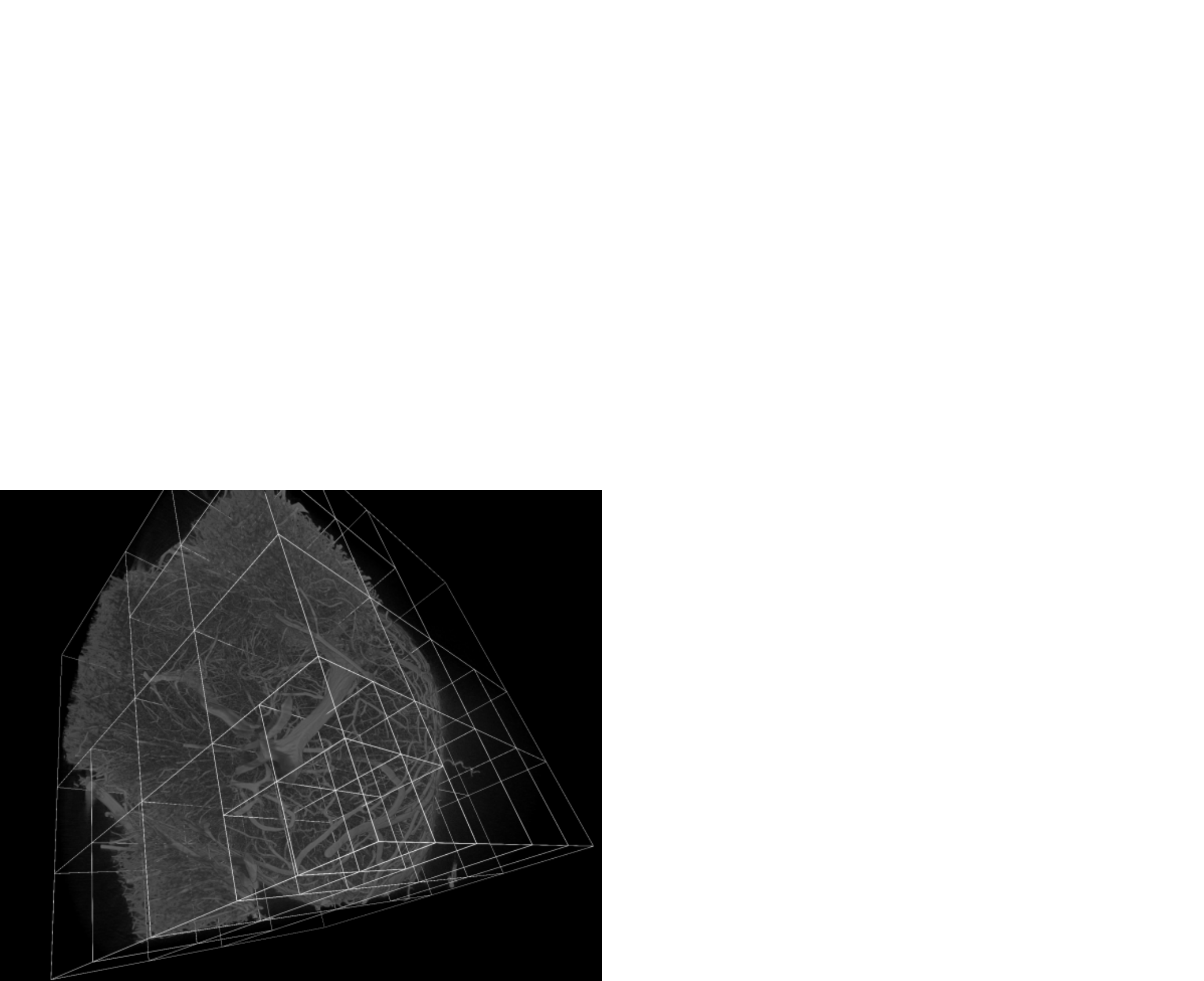}}\\
\begin{tabular}{ccc}
\subfloat[]{\includegraphics[width=1.0in]{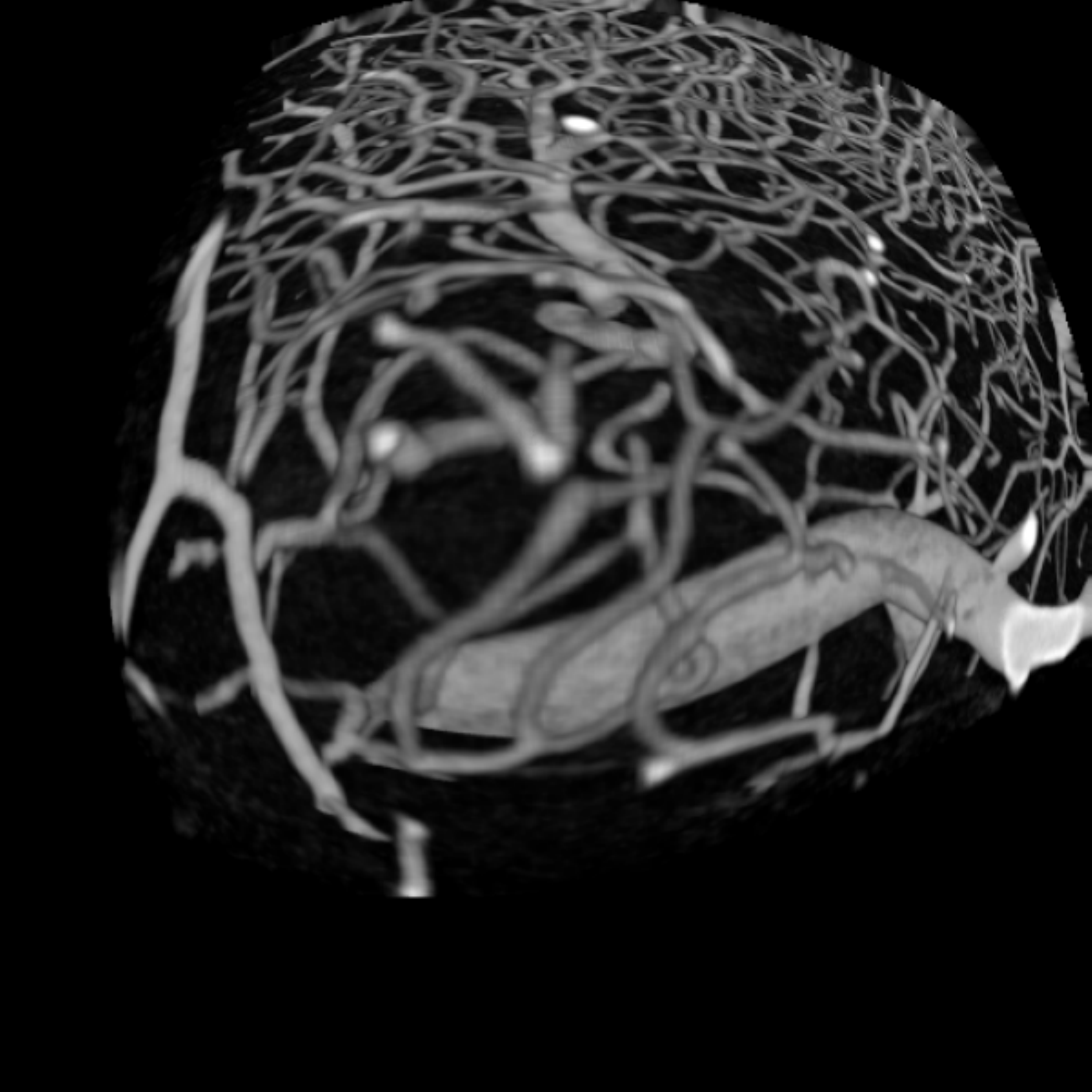}} & \subfloat[]{\includegraphics[width=1.0in]{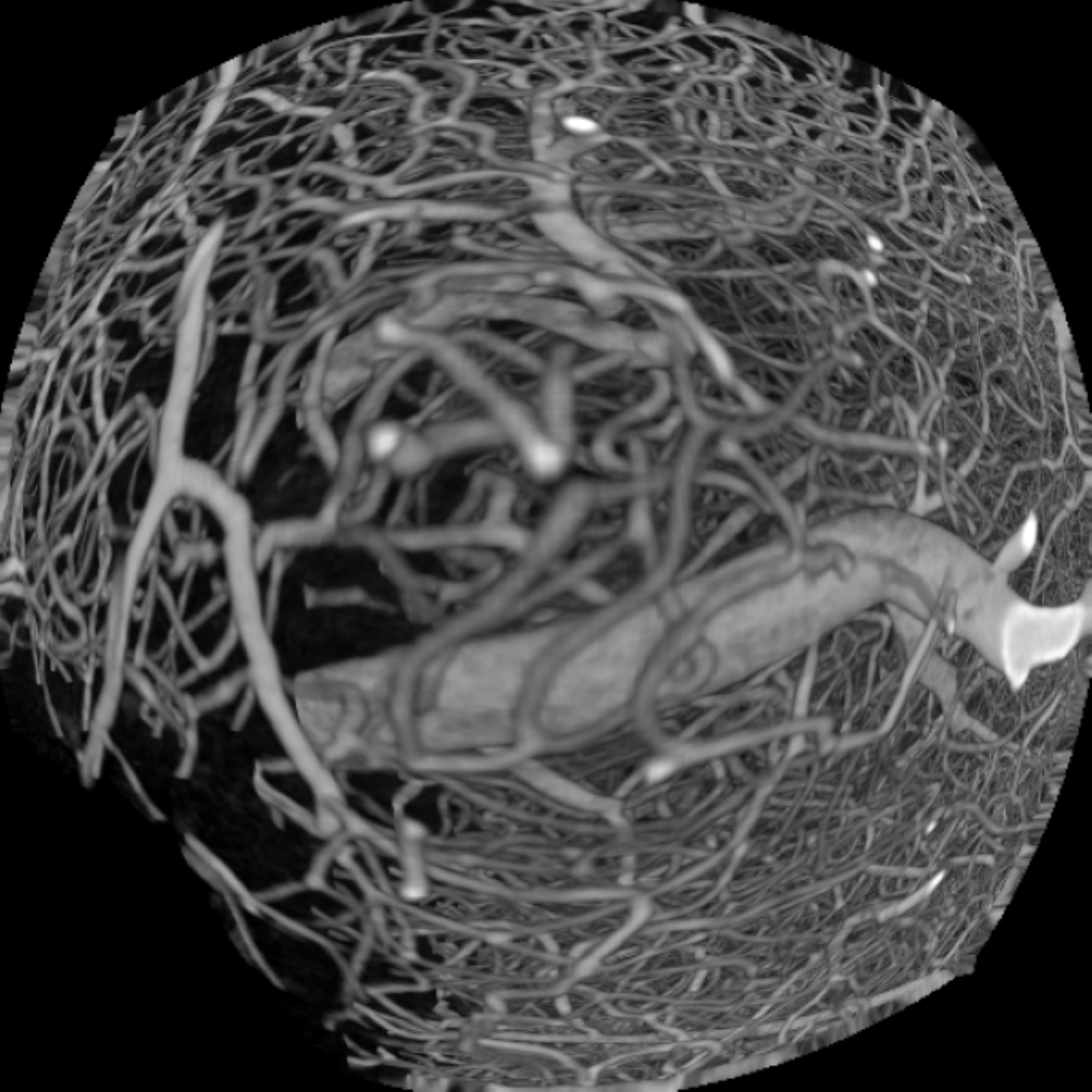}} & \subfloat[]{\includegraphics[width=1.0in]{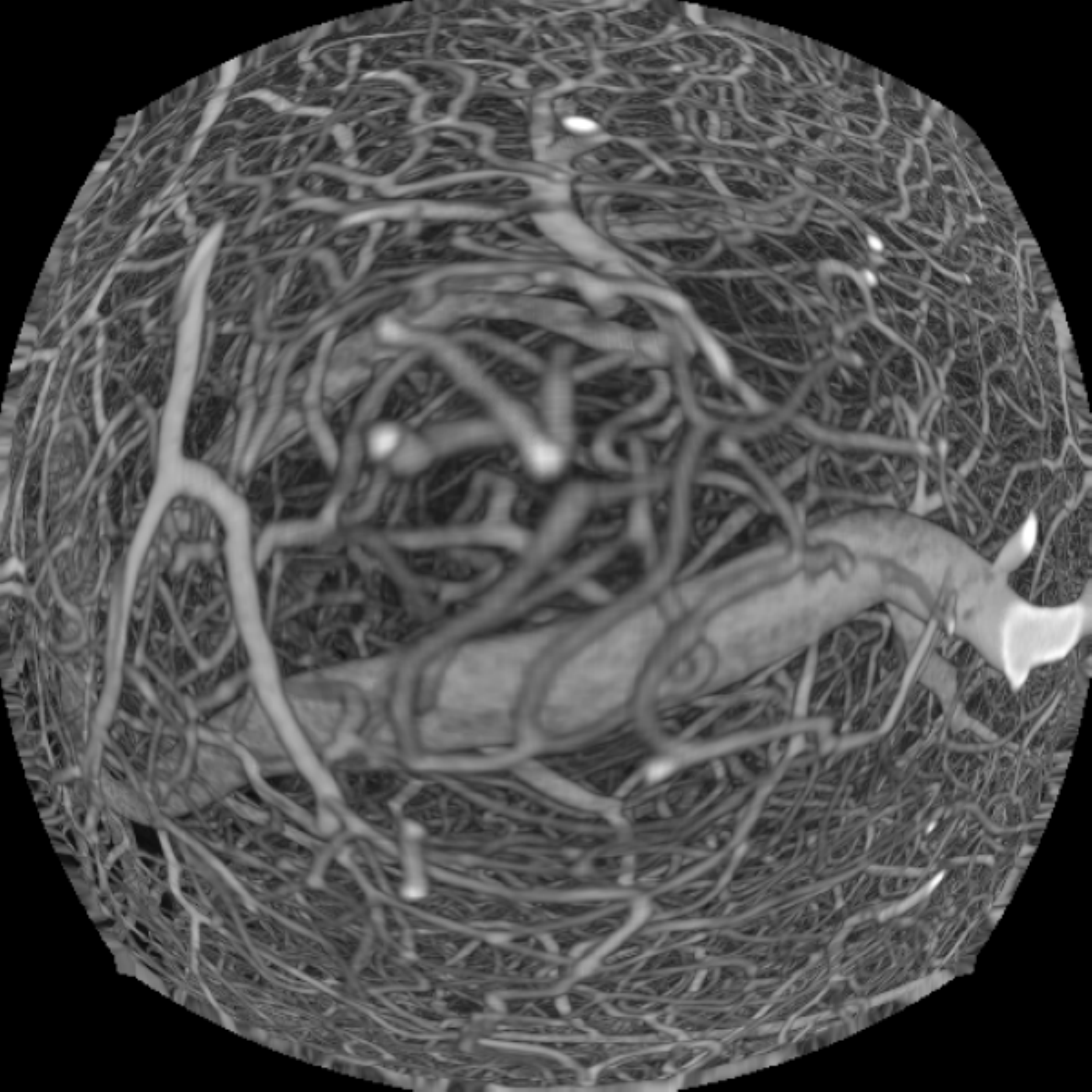}} \\
\end{tabular} 
\subfloat[]{\includegraphics[width=2.5in]{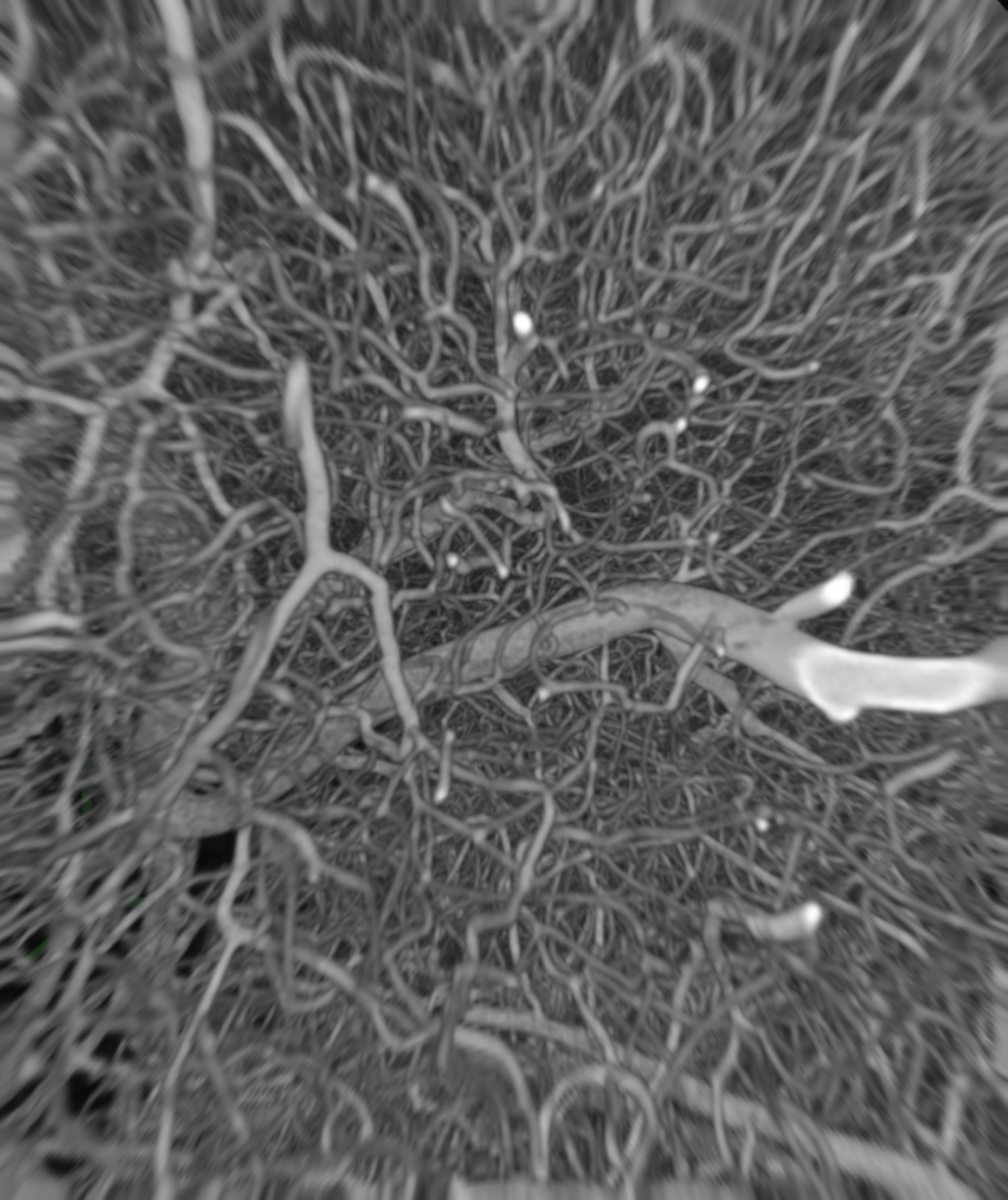}}
\caption{\textbf{Octree multipass rendering.} a) Volumetric data where octree blocks in the render-list are marked with white bounding boxes; b) Result of the first pass in wrapped space; c) Accumulated image after half of all passes were completed; d) Result after all passes were done; e) Unwrapped image. Data courtesy by Dr. James Simpkins, West Virginia University, representing a mouse brain vasculature imaged via CAT.}
\label{fig:octree_multipass}
\end{figure}

We render octree blocks with a multipass approach. The octree traversal happens on CPU, during which a schedule of blocks to be rendered is created. Each block is rendered using a separate draw call in one pass. All the blocks are sorted and rendered from the front to the back block, which guarantees correct integration for the emission-absorption model, and allows early ray termination. All the blocks are rendered to the intermediate framebuffer with color and alpha channels in a wrapped 2D space. When a subsequent block is rendered, the ray is continued inside that block by reading previously written values of color and alpha. After the ray reaches the end of the current block or has terminated, new values of color and alpha are written back. See Figure~\ref{fig:octree_multipass}.

% This entire paragraph could be eliminated entirely if we need space
The octree traversal algorithm starts from the root node and then checks if the current resolution level is enough or no. If it suits, then the current block is registered in the render list. If not, the procedure repeats recursively for all children nodes. Children nodes are traversed in the order that matches the overlapping order, thus the resulting render list appears to be sorted from the front to the back block. The traversal algorithm also performs frustum culling of the blocks by ignoring not visible ones. The decision on whether the current resolution level suits the needs of creating a quality rendering for the user or not is taken by comparing the needed angular voxel resolution with the angular resolution provided by the closest face of the block at the given distance. If one or more of the children blocks are not in the cache, they are added to the request queue and one level above resolution is used. When all children blocks are loaded, the needed resolution level is used.

% \begin{figure}[th!]
% \centering
% \includegraphics[height=2.0in]{20170611022854}
% \caption{\textbf{Octree based visualization.} Leafs of the octree are marked with white bounding boxes.}
% \label{fig:octree}
% \end{figure}

\subsection{DVR for HMD}

Current HMD technology requires lenses to accommodate the installation of displays close to the eyes. This comes at the expense of introducing pronounced pincushion distortion. HMD drivers accept input images in the unwrapped space to wrap them immediately thereafter to compensate for the distortion. The wrapping operation causes information loss because the input image is downsampled. Therefore, rather than investing DVR computing power for rendering information that will not be used, we have increased the efficiency of the ray casting process by rendering the volume directly in the wrapped space, with uniform ray distribution across the wrapped image. Then, the image is unwrapped to the original space, where the ray distribution becomes non uniform. This technique requires a significantly lower amount of rays to be cast, and improves performance without any noticeable image quality degradation.

\subsection{3D Movies}

In syGlass we have implemented a playback functionality for rendering in VR sequences of volumetric data frames, or 3D movies. Movies with volume frames made up to $20\times10^6$ voxels are played at  high volume frame rate (25FPS). That is achieved thanks to our volumetric data container architecture that is optimized for data transfer and visualization operations.

\subsection{Annotation Data Rendering}

Annotation data in VR can vary rapidly over time because of changes made by the
user, and there can be a large amount of geometric primitives for annotations in
one scene, which typically include dots, lines, spheres, cylinders and cones.
For example, for skeletonizing a biological structure like a neuron, one can use
a graph where each node has attributes like position, radius and color. Such a skeleton graph would be rendered with nodes given by spheres with the attribute radius, and edges made by cones tangent to the spheres. Generation of meshes for annotation data on CPU may not be an optimal solution since this data requires frequent updates. Besides, rendering smooth surfaces will require a significant amount of polygons per sphere and edge, that can quickly create a bottleneck in vertex processing.
\begin{figure}[t!]
\centering
\begin{tabular}{cc}
\subfloat[]{\includegraphics[width=1.5in]{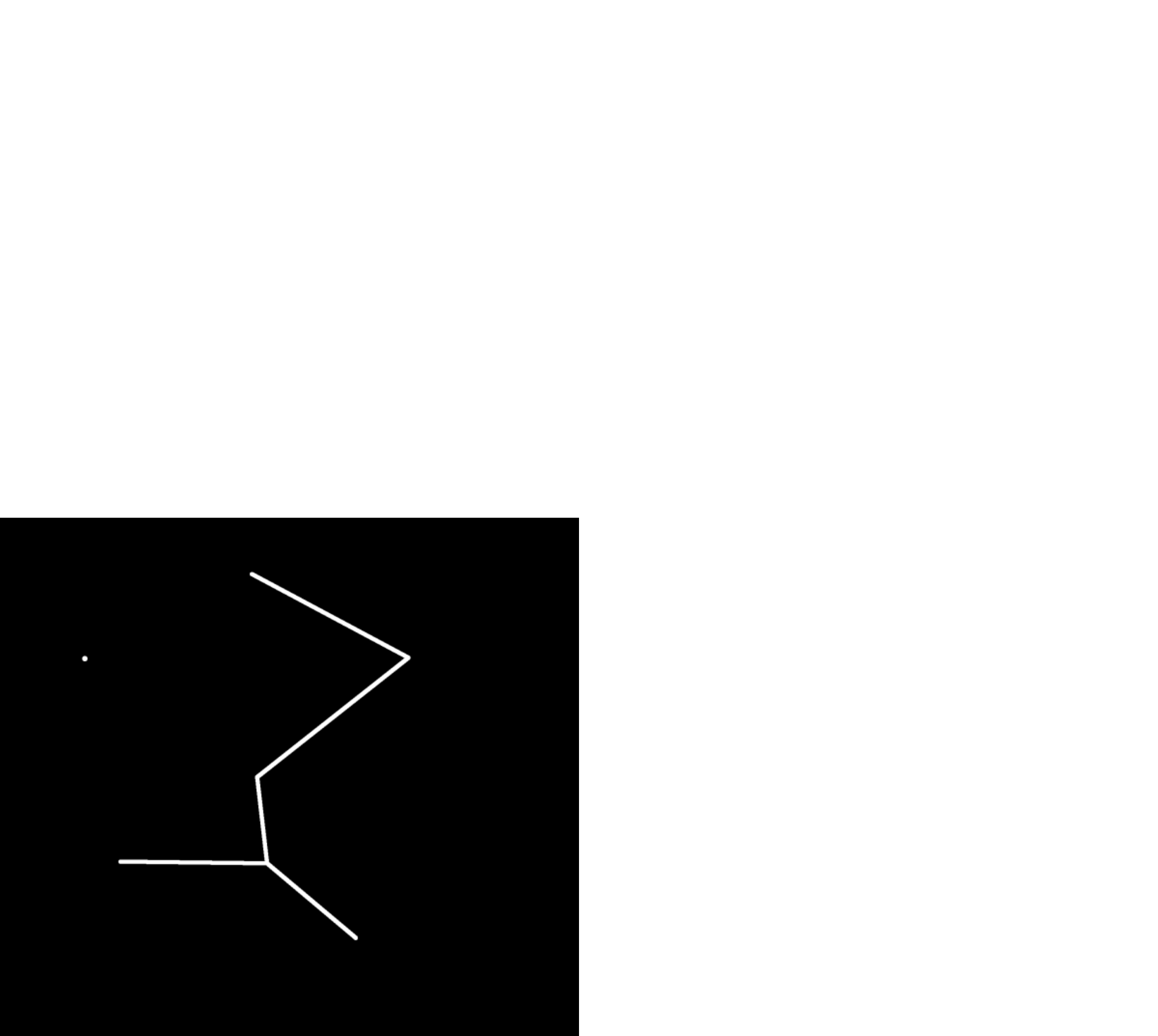}} & \subfloat[]{\includegraphics[width=1.5in]{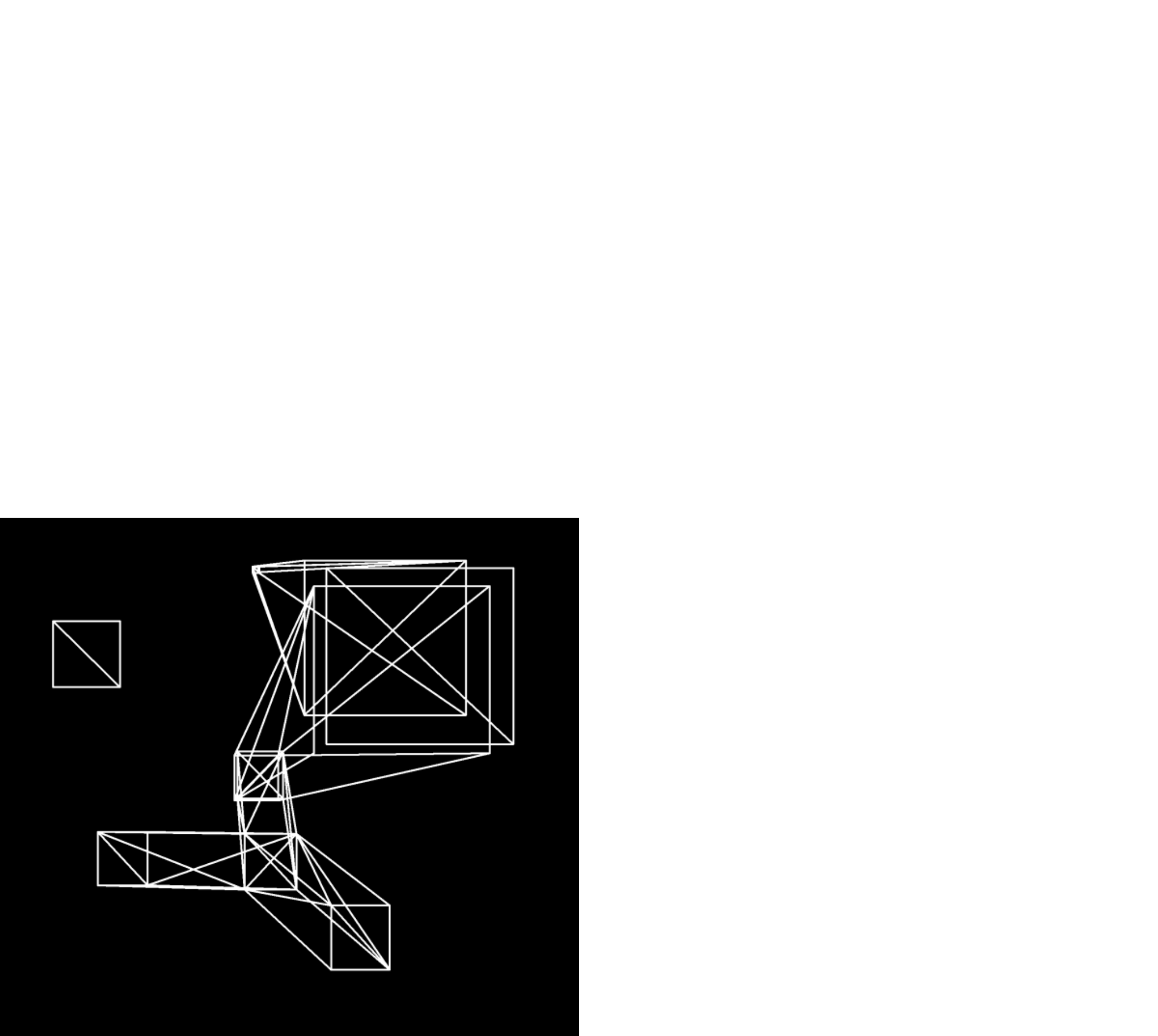}}  \\
\end{tabular} 
\subfloat[]{\includegraphics[width=1.5in]{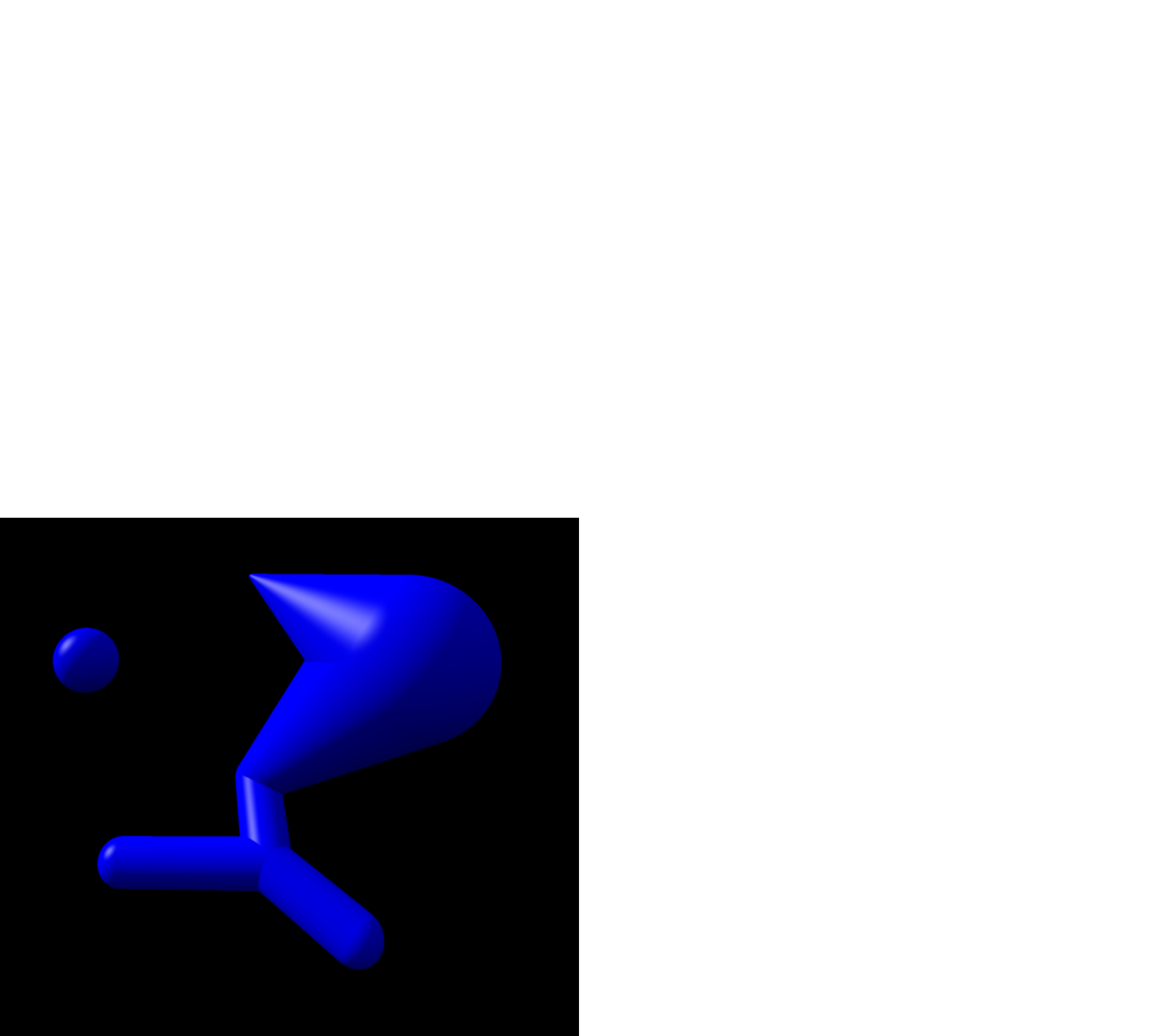}}
\caption{\textbf{Annotation data rendering with imposters.} a) Input "guide" geometry transformed by the vertex shader; b) Primitives emitted by the geometry shader; c) Result of the fragment shader.}
\label{fig:imposters}
\end{figure}
A similar problem appears in the visualization of molecular dynamics simulations. In that field, imposters were proven to be a very efficient solution~\cite{Zwan,doi:10.1093/bioinformatics/btu426}. 
% In molecular dynamics visualization systems, imposters are used to render a large amount of spheres and cylinders. We extend the same ideas to render tangent cones that connect two spheres of different diameters.

Imposters are planes which are rendered faced to the viewer, and have a texture that depicts the object of interest. Imposters should contain the least possible amount of geometry and cover the represented object in screenspace tightly. %, meaning that it should have minimal, but sufficient area to fully overlap the object in screenspace. 
The texture is generated on-the-fly in the fragment shader using ray casting. The fragment shader performs shading of the object using analytically calculated normals and invalidates pixels that do not belong to the projection of the object. Imposter geometry can be generated on-the-fly directly on GPU out of ``guide'' geometry. This represents a minimal set of primitives with attributes, given by bounding planes computed by the geometry shader, which is an optional shader pass that follows the vertex shader and preceded the fragment shader.

We render annotation data with imposters. This allows to render annotation
(geometric) primitives using ray casting, and to have minimal impact on memory
usage because only guide geometry needs to be maintained.
Figure~\ref{fig:imposters} shows the input, given by the guide geometry, the
primitives emitted by the geometry shader and the final result produced by ray casting and shading in the fragment shader.

% Imposters allow:
% \begin{itemize}
%   \item  Rendering of geometric primitives using raycasting, which gives smooth, analytical surfaces
%   \item  Minimal impact on memory usage. In memory stored only "guide" geometry
%   \item  Fast update of the scene, since only "guide" geometry should be updated.
% \end{itemize}

\subsection{Volumetric and Mesh Data Compositing}

\begin{figure}[t!]
\begin{tabular}{cc}
\subfloat[]{\includegraphics[width=1.6in]{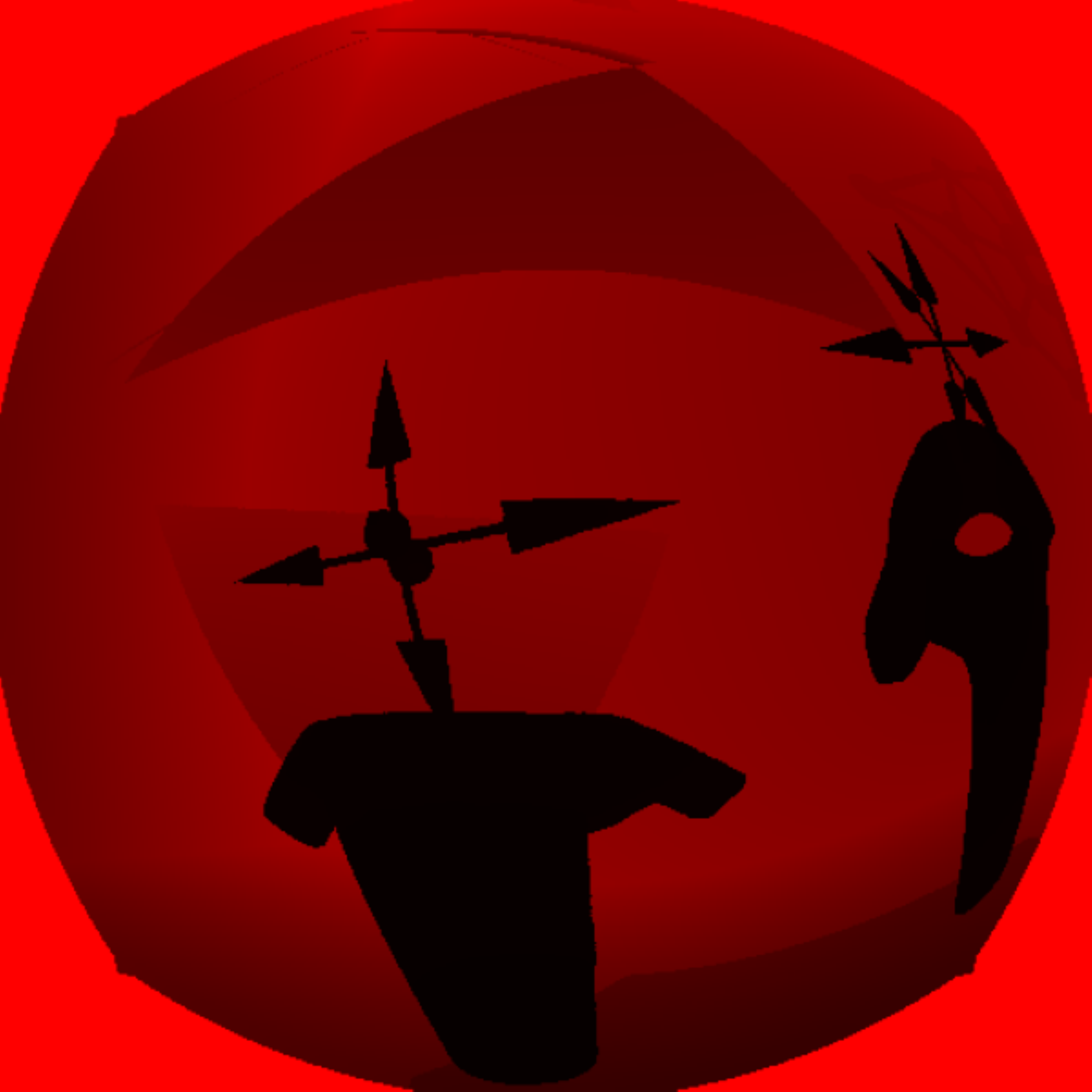}} & \subfloat[]{\includegraphics[width=1.6in]{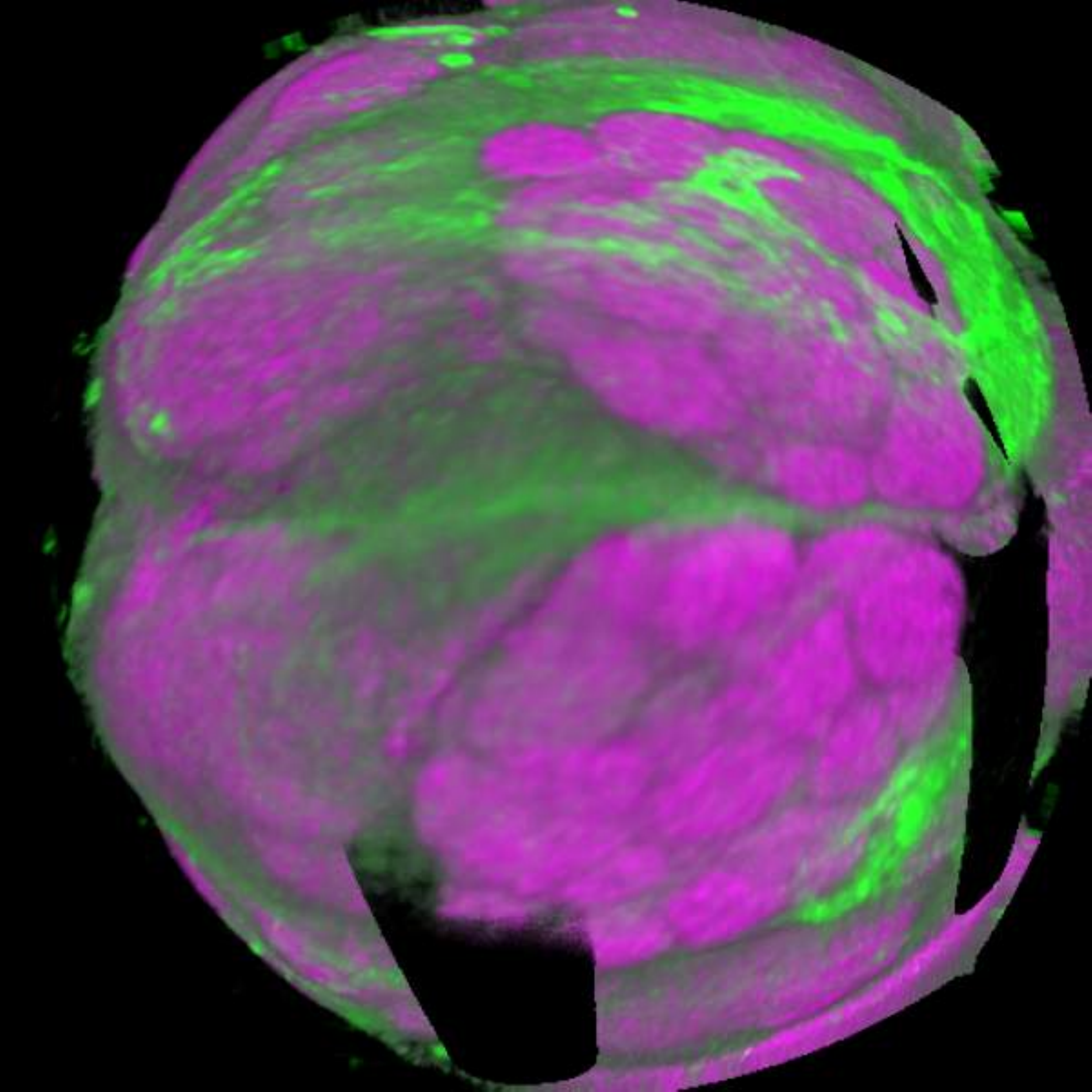}} \\
\subfloat[]{\includegraphics[width=1.6in]{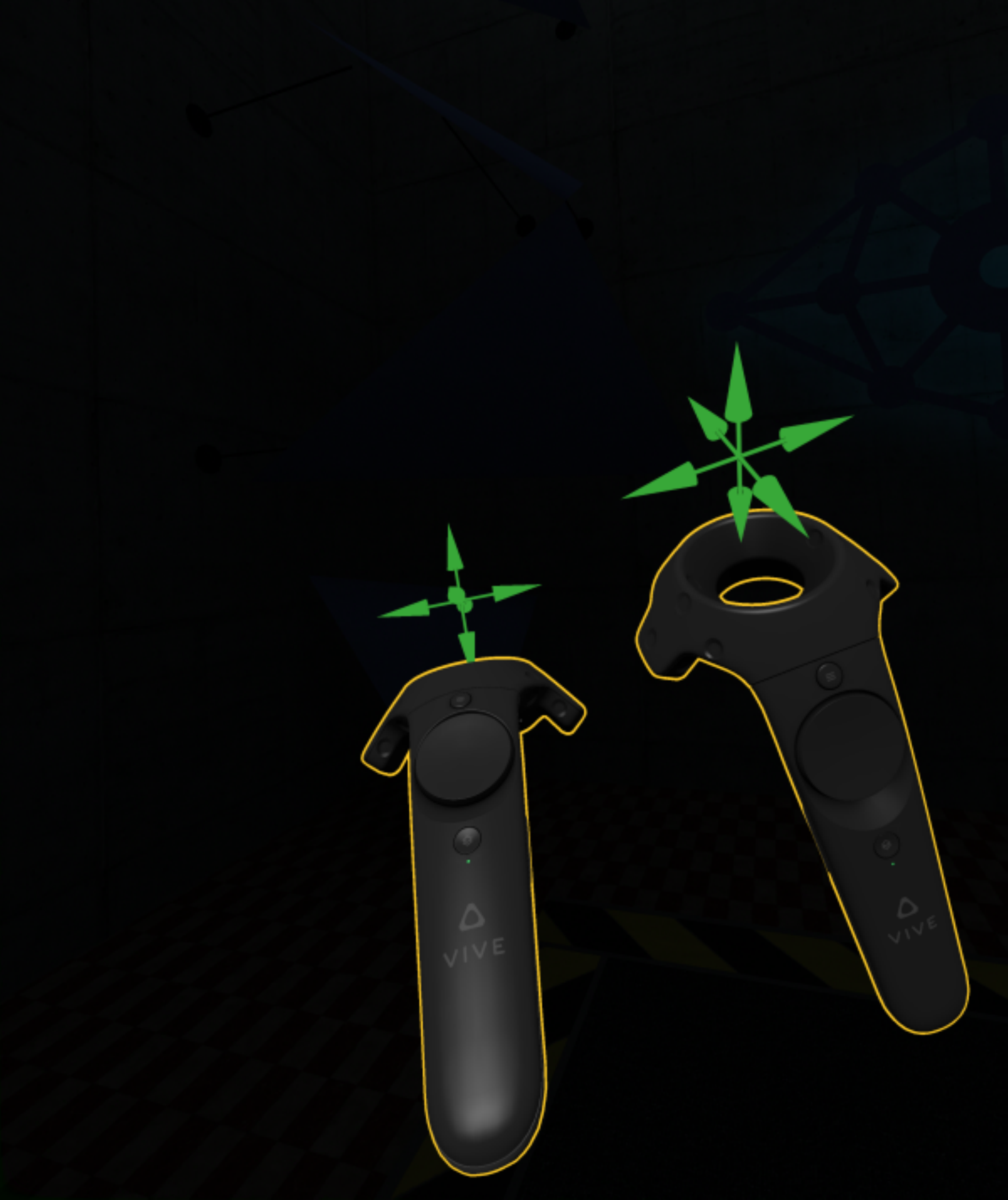}} & \subfloat[]{\includegraphics[width=1.6in]{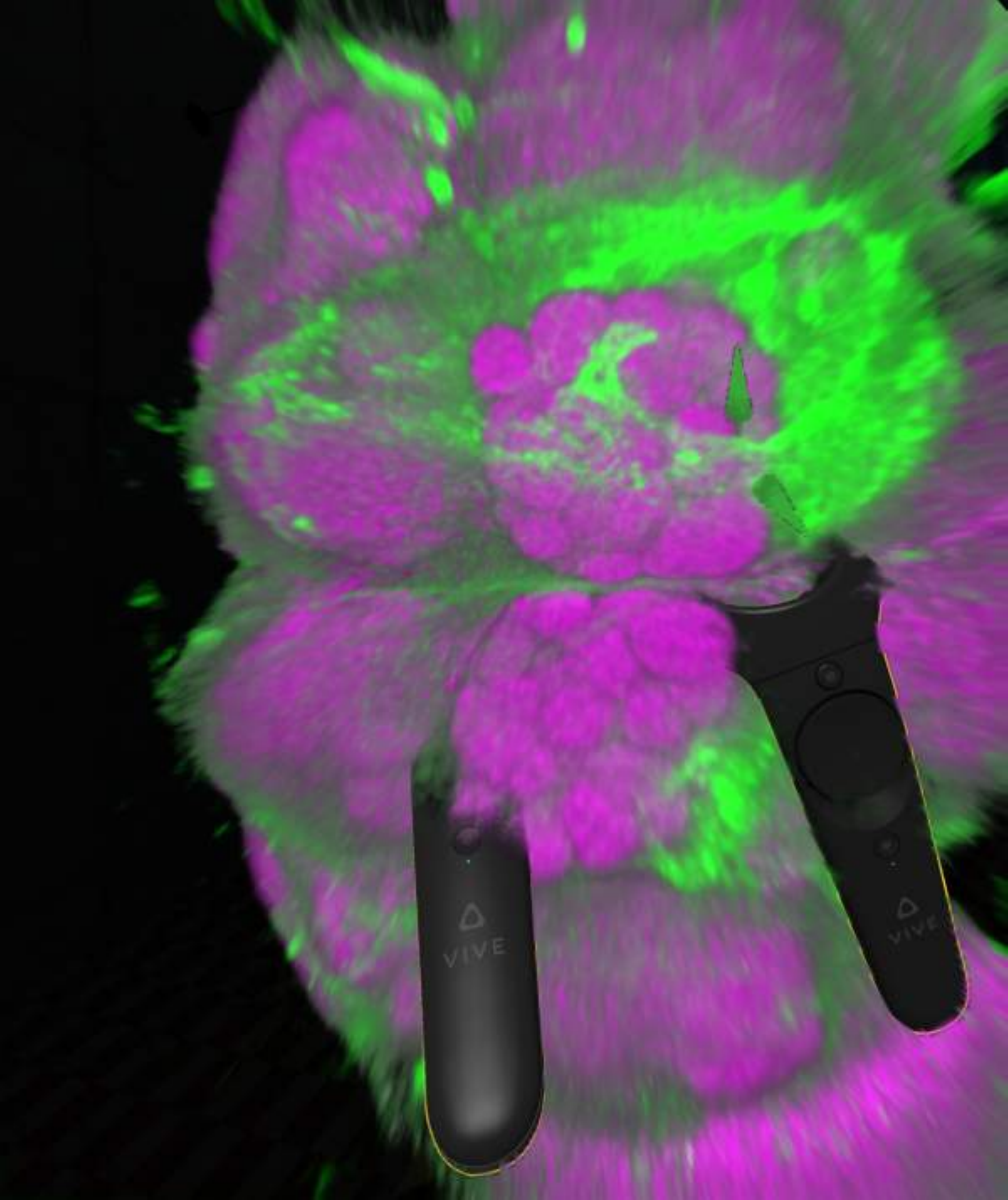}} \\
\end{tabular} 
\subfloat[]{\includegraphics[width=3.5in]{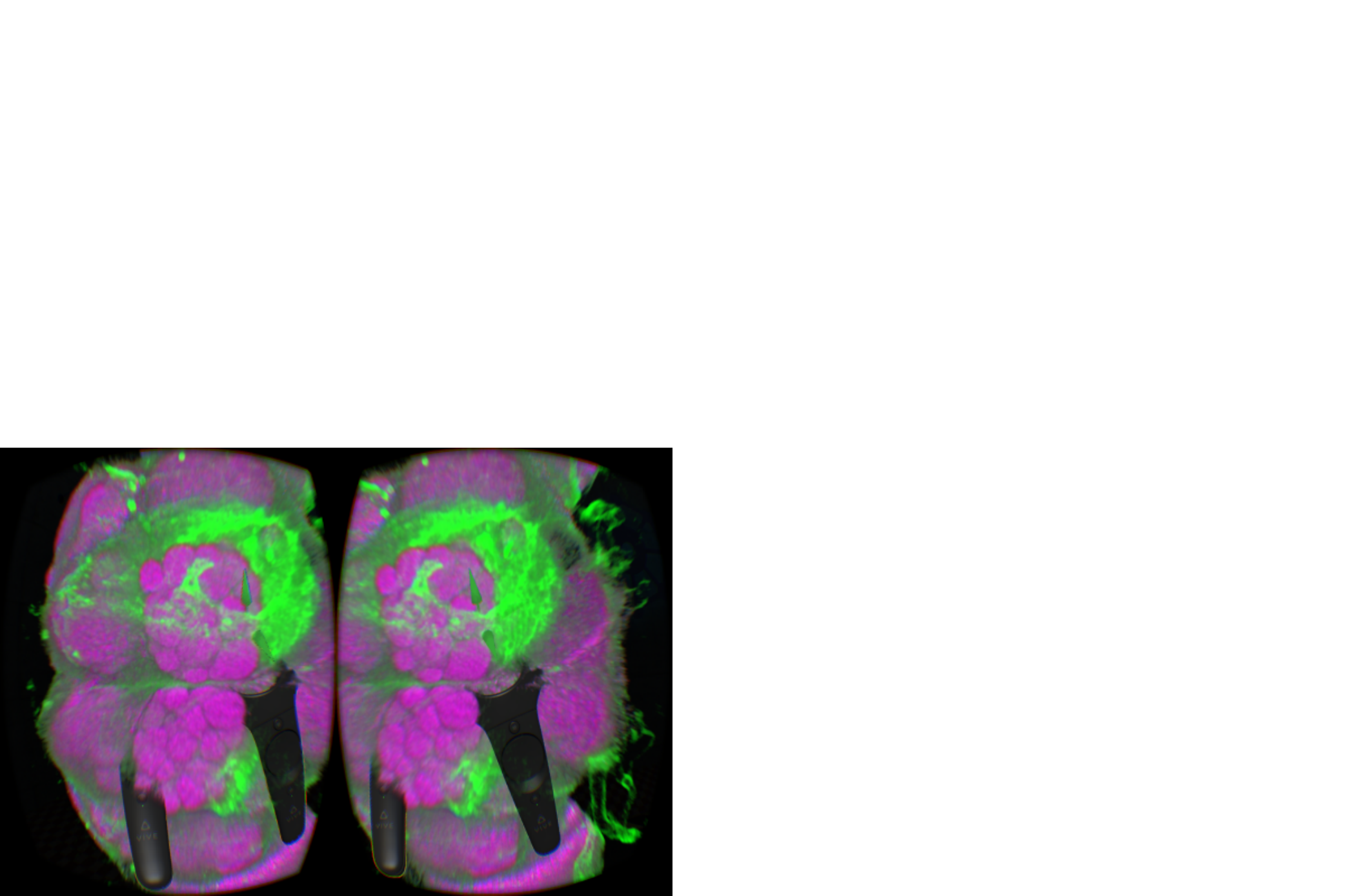}}
\caption{\textbf{Volume data compositing.} a) Wrapped depth buffer; b) DVR buffer in wrapped space, computed with depth buffer ray termination; c) Previously rendered mesh scene, which was the source of depth information; d) Mesh scene combined with the unwrapped DVR buffer; e) Final side-by-side image with applied Barrel distortion to compensate for the lenses distortion.}
\label{fig:compositing}
\end{figure}

If we consider the volume as occupied by volumetric data in front of mesh or annotation data, we can proceed by ray casting volumetric data, and then overlay the projection on top of the scene behind. When the emission-only model is used, additive blending is the natural way for combining the two images, since volume does not absorb light and should not reduce the intensity of the geometry behind. When the emission-absorption model is used, alpha compositing would be the proper choice.

Images produced by DVR always have an alpha channel with the accumulated alpha value along the ray. The emission-only model outputs the computed color with a zero alpha, while the emission-absorption model can output the color pre-multiplied by the alpha value. Therefore, the two models can be handled seamlessly using pre-multiplied alpha compositing.

The strategy above clearly breaks down when not all volumetric data is in front.
Indeed, annotation data, often found in front, would be rendered in a scene where stereoscopic vision would not allow the correct depth perception of both types of data, making it hard to use annotation tools.

To render a geometric scene in combination with volumetric data, we begin by
rendering the scene and saving the depth buffer into a texture. Then, the
content of the depth texture is rendered to a second framebuffer and saved onto
a float-point texture, which now contains the actual distances from the camera
to the fragments, and is also wrapped to match the wrapped space in which the
volume is rendered. During ray casting, a given ray is terminated at the
distance stored in the corresponding position in the float-point texture, and
thus treated as a ray that has encountered an opaque surface. As the last step, pre-multiplied alpha compositing is used for image blending, as described earlier. Figure~\ref{fig:compositing} provides an illustration of the steps of this compositing strategy.

%%% Local Variables:
%%% mode: latex
%%% TeX-master: "syglasspaper-main"
%%% End:

\section{Analytical Tools}\label{annotation_tools}

In syGlass we have developed a number of tools for enabling the user to adequately interact with raw and annotation data effectively enabling analysis in VR while wearing a HMD. To do so we had to render in VR the HMD controllers in a way suitable for an adequate UI supporting the desired type of analysis. The tools operate on data represented in a metric space, so that measurements reflect the real geometric properties of the tissue visualized. The tools are made available to the user by a GUI developed in VR that allows their selection by pointing towards the associated icon. The specific analytical tools developed are briefly described below. 

\begin{figure}[t!]
  \begin{tabular}{ccc}
\centering
    \subfloat[] {\includegraphics[width=1.0in]{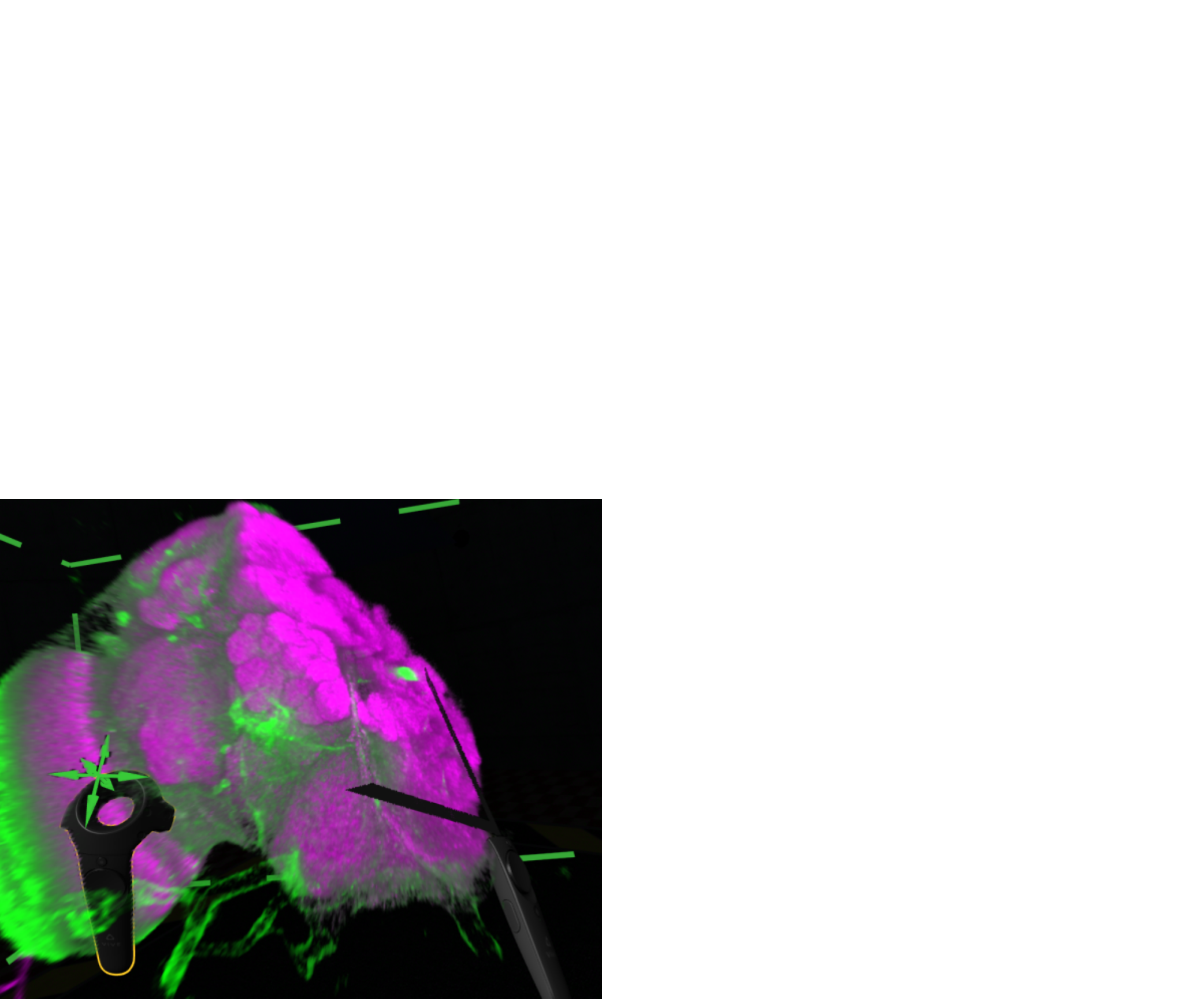}\label{transform_cut}} &
    \subfloat[] {\includegraphics[width=1.0in]{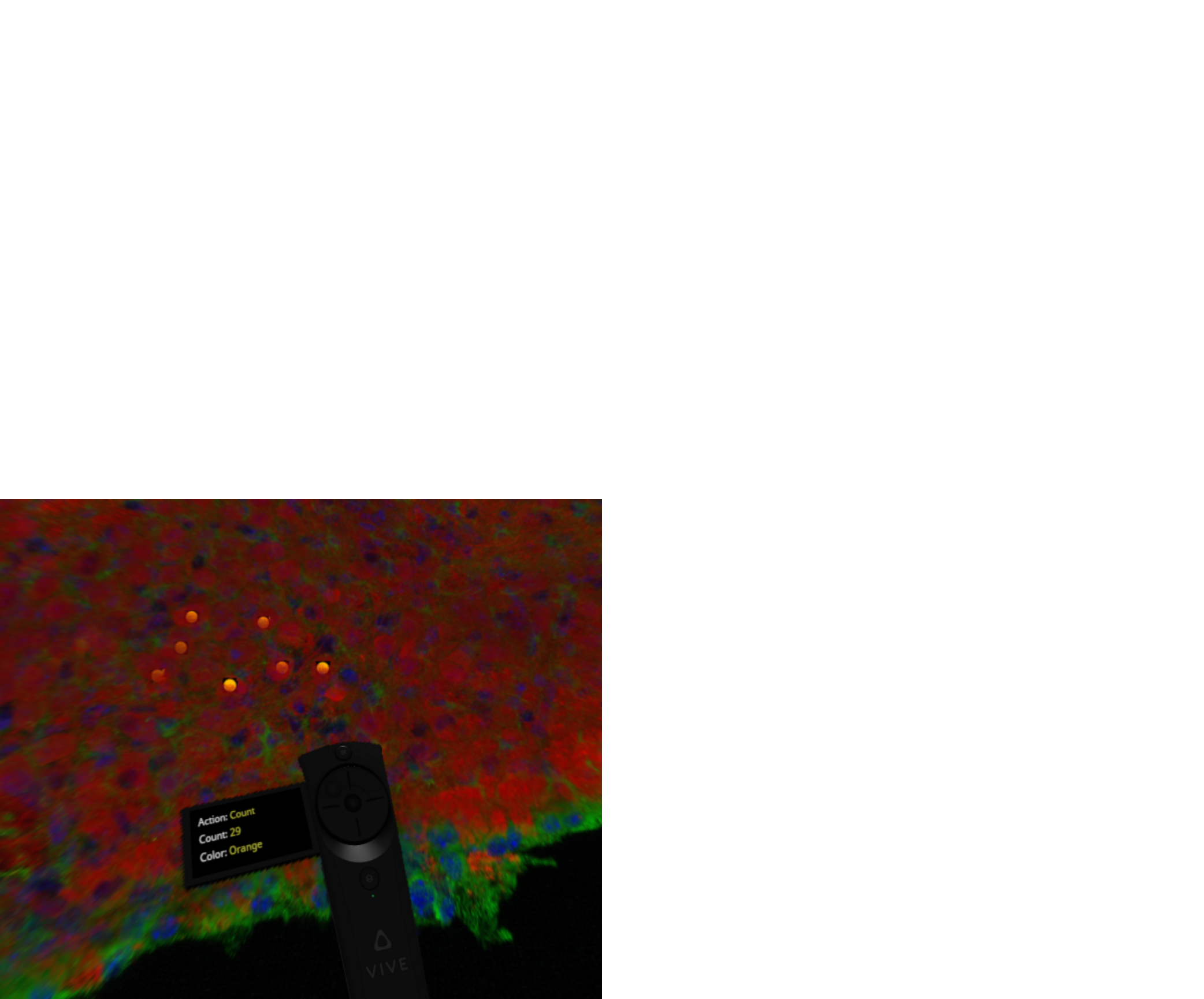}\label{counting}} &
    \subfloat[] {\includegraphics[width=1.0in]{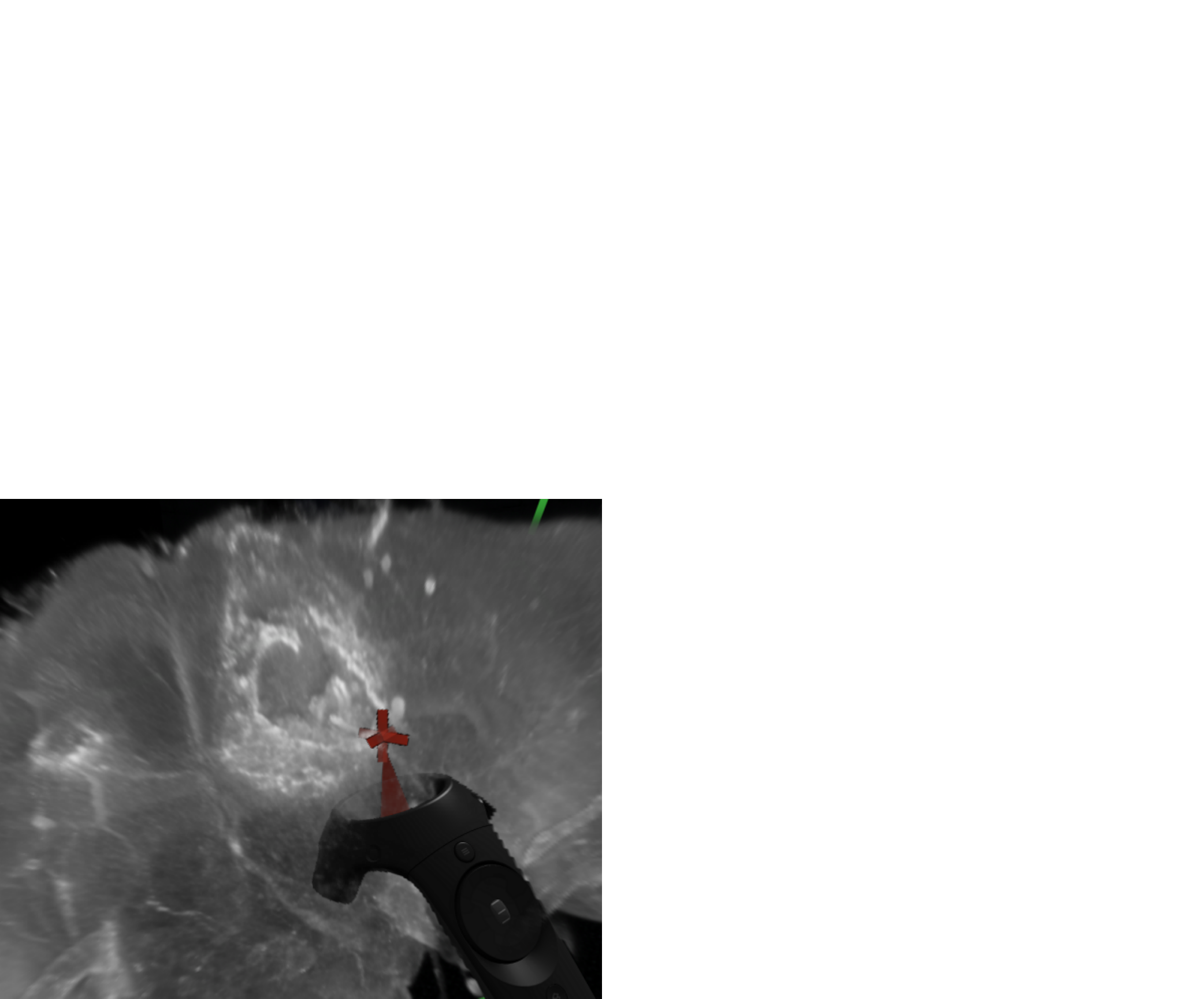}\label{camera}} \\
  \end{tabular}
  \begin{tabular}{cc}
    \subfloat[] {\includegraphics[width=1.59in]{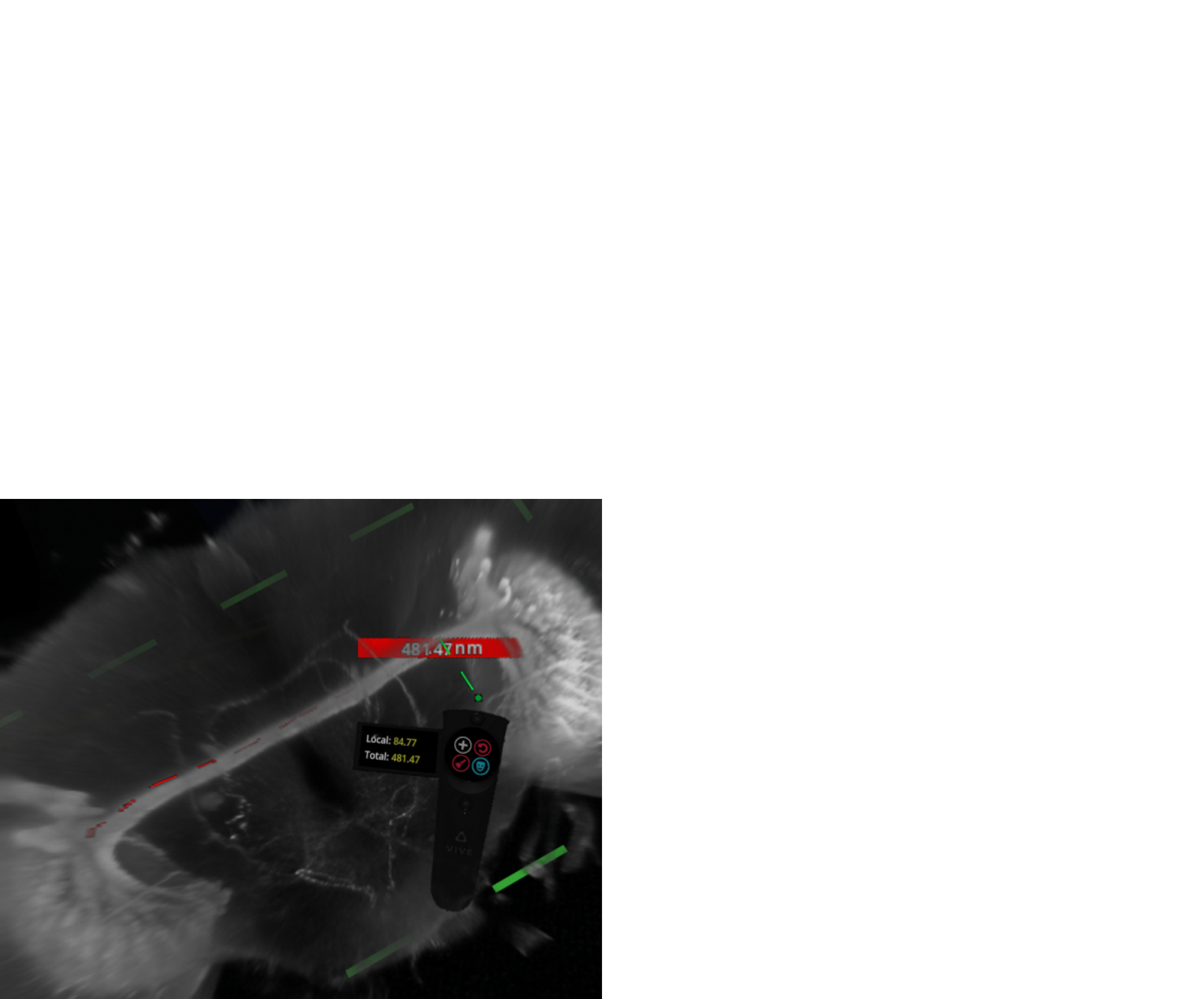}\label{measuring}} &
    \subfloat[] {\includegraphics[width=1.59in]{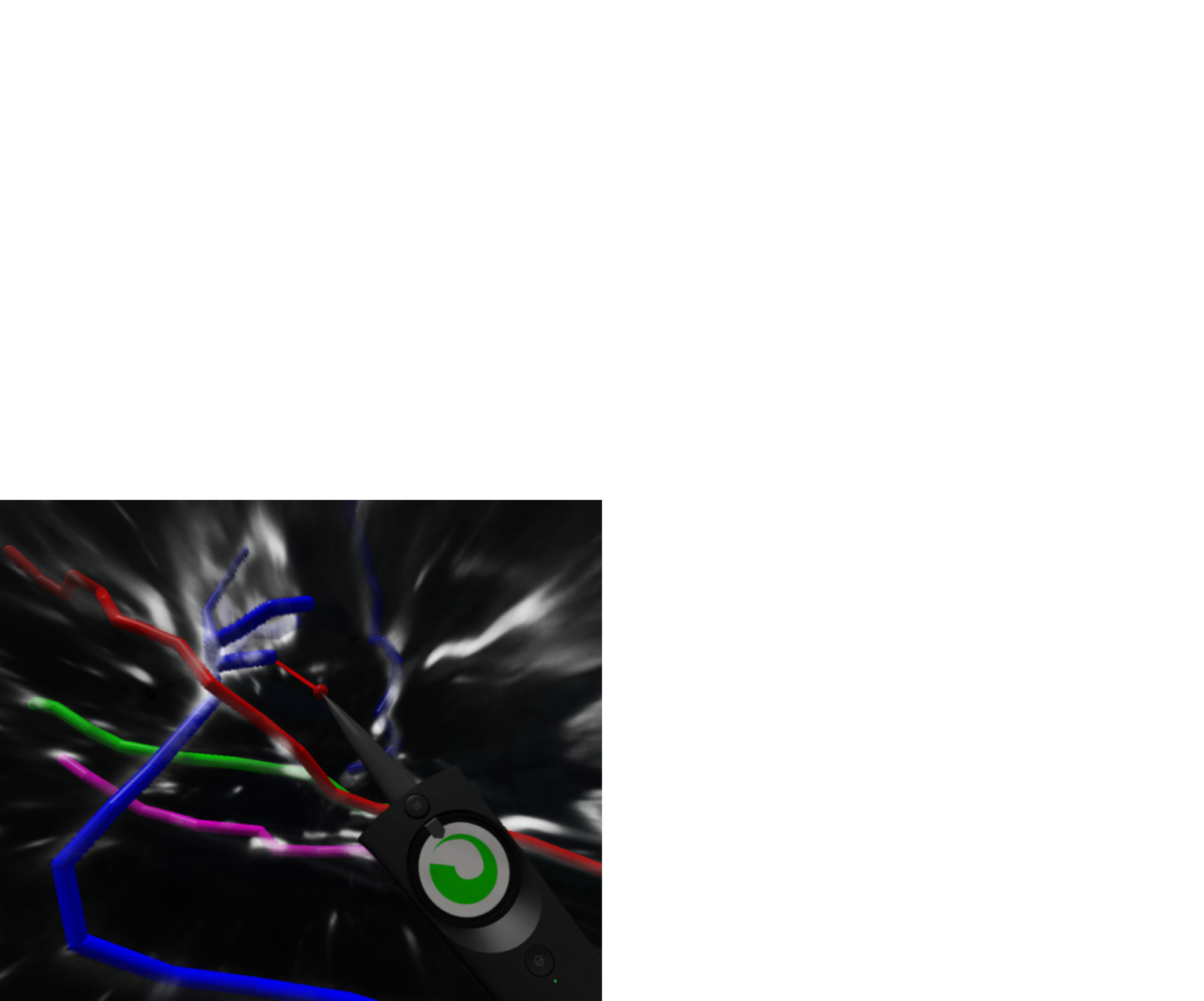}\label{skeleton}} \\
    \subfloat[] {\includegraphics[width=1.59in]{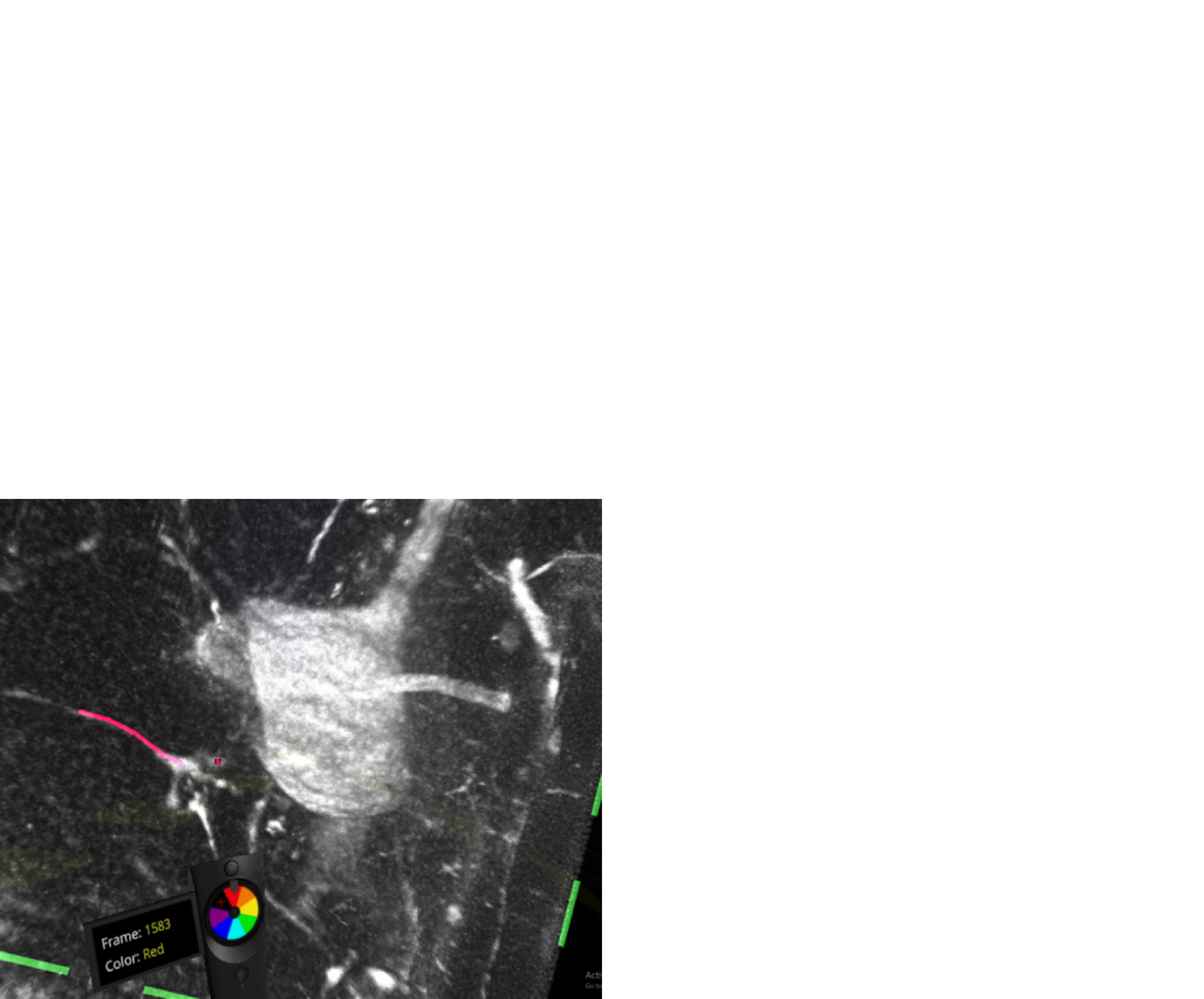}\label{tracking}} &
    
    \subfloat[] {\includegraphics[width=1.59in]{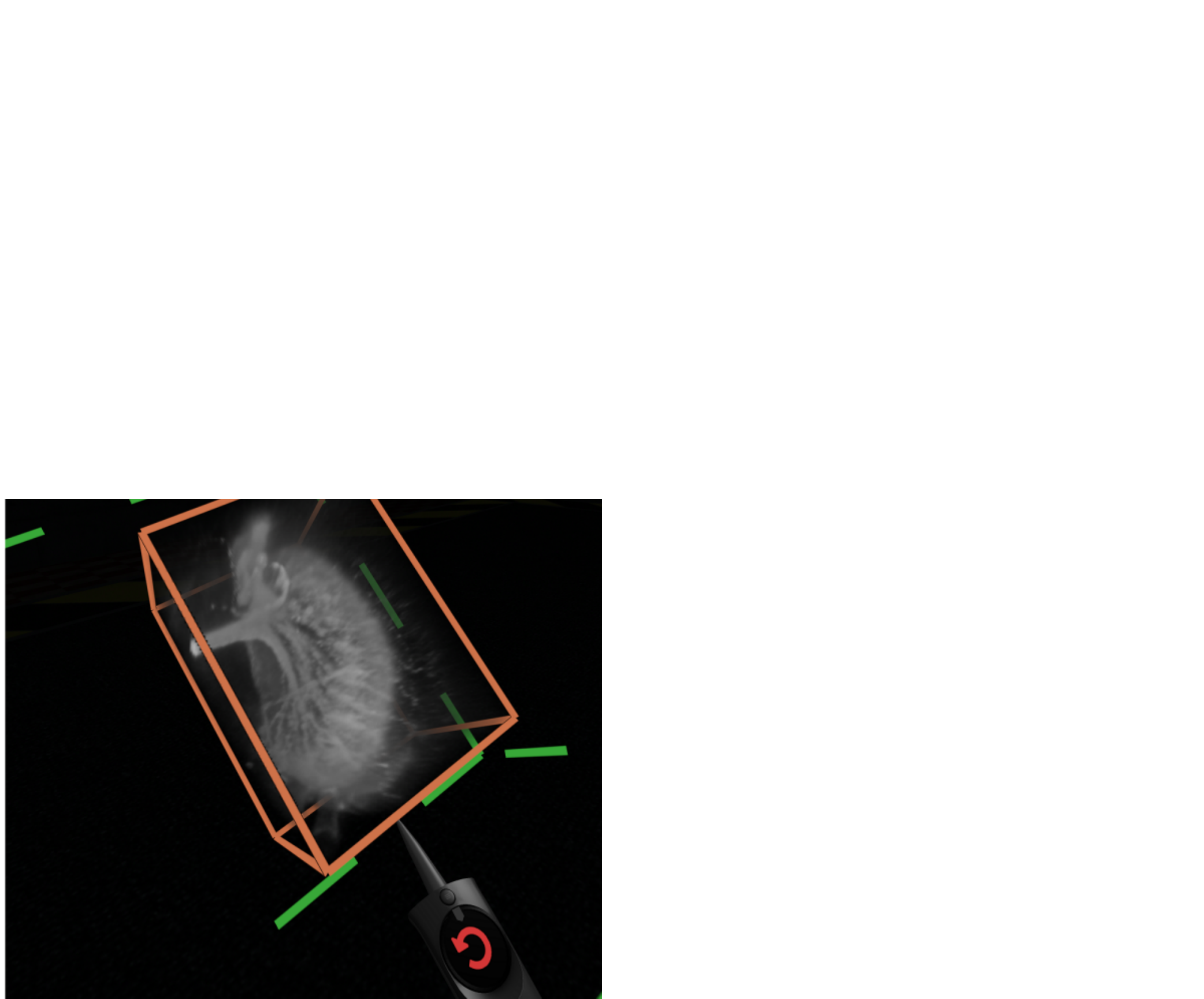}\label{roi}} \\
  \end{tabular}

  \caption{\textbf{Tools in action.} a) The transformation and cut plane tools; b) The counting tool; c) The camera tool; d) The measuring tool; e) The skeleton tracing tool; f) The video tracking tool; g) The region of interest tool.}
\label{fig:tools}
\end{figure}

\textbf{Transformation tool:} The transformation tool allows the rotation, translation, and scaling of a data volume interactively in VR by the user grabbing the data with the controllers to perform this interaction. The same tool is also able to select a new viewer position inside the virtual room where the data is presented. This functionality corresponds to a form of tele-transport inside the VR environment. See Figure~\ref{fig:tools}\subref{transform_cut}

\textbf{Cut plane tool:} The cut plane tool transforms a hand-held controller into a virtual scissors capable of instantly cutting a volume along a plane, and see the volume as if it was missing the portion above such plane. See Figure~\ref{fig:tools}\subref{transform_cut}. This functionality operates for both volumetric and mesh type of raw data, and extends also to the visualization of the original image slices composing the raw image stack of the project.

\textbf{Counting tool:} This tool allows to count objects in VR. It allows the user to manipulate the controller and virtually touch the object to be counted while simultaneously placing a virtual sphere or ball, which also marks that the object has been counted. The balls can be colored differently to identify
different types of objects being counted, and syGlass automatically counts the number of balls placed. See Figure~\ref{fig:tools}\subref{counting}.

\textbf{Camera tool:} The camera tool is used during cataloging and the
production of cards that are stored in the Annotation Database, and are managed
by syBook. In VR mode, the user can use this tool and capture a snapshot of the
scene which is stored inside a new card that is added to the database as
discussed further in Section~\ref{cards_section}. See Figure~\ref{fig:tools}\subref{camera}. In addition, other information describing the status of syGlass are also recorded in the card. Besides taking snapshots, the camera tool is also able to shoot a movie corresponding to a predefined trajectory of the viewer.

\textbf{Measuring tool:} This tool is capable of measuring the distance between arbitrary points on the volume. When a project is setup initially, the user adds the dimensions of the data, with this information the system sets up the metric space which allows the system to report true distance, as can be seen in Figure~\ref{fig:tools}\subref{measuring}.

\textbf{Skeleton tracing tool:} The skeleton tracing tool allows the user to use the controller in VR and to build a tree-like structure in 3D. In particular, the tree branches can have different thickness, bifurcations can have arbitrary orders, and every tree component can be edited interactively in VR. The tool can be used for tracing the skeleton of complex three dimensional biological structures like neuronal cells. See Figure~\ref{fig:tools}\subref{skeleton}. The information defining such tree structures can be exported onto a file format called SWC, often used by neuroscientists. syGlass can also load raw data, and an associated SWC file and render them simultaneously in VR, and provide the ability to edit the tree structure.

\textbf{Feature tracking tool:} syGlass is capable of playing volumetric movies in VR, characterized by a temporal sequence of 3D volumes. The feature tracking tool allows the user to track a volumetric feature point over time by placing markers in the virtual space, and by doing so over time, thus creating a linked temporal track as the feature moves. See Figure~\ref{fig:tools}\subref{tracking}.

\textbf{ROI tool:} The region of interest (ROI) tool allows to select a rectangular cuboidal region of the virtual space. See Figure~\ref{fig:tools}\subref{roi}. This region could then be used to pose queries to the Annotation Database, or to cut out of the rendering pipeline the data outside of the ROI volume.

\section{Annotation Cards}\label{cards_section}

We found that users often need to have a way to save their place in the work they were doing with a volume. Not only so that they could pause work, such as annotation or their investigation, but also in the case that the user would like to catalog and share a finding with a colleague, or where annotation work may be a collaborative effort.  For this case, we designed a facet of the system which we refer to as \textit{cards}. Cards have a number of features which allow for the efficient return to a viewpoint on a volume or set of meshes.

Each card can carry a variety of meta-information to describe the reasoning behind creating the card including: voice memos, text descriptions which can be searched for card retrieval, and an image rendered through the user's viewpoint at the time of creation.

Cards also act as a state serialization mechanism, which stores and refreshes: the shader settings currently in use, the user's position relative to the data, and the data's exact position in the VR space.

The user can browse cards with an interface in syBook. The user can view all of the cards they have access to in various views such as a grid or a timeline view, or users can view cards as grouped under the projects they relate to. Once a card is interacted with, the user is presented with all the
meta-information stored on that card, and she is given the opportunity to edit those details. Finally, a launch button allows the user to boot the VR system directly into the state stored by the card.

%%% Local Variables:
%%% mode: latex
%%% TeX-master: "syglasspaper-main"
%%% End:

\section{Visualization Enhancements}

\begin{figure}[t!]
\centering
\begin{tabular}{cc}
\subfloat[]{\includegraphics[width=1.5in]{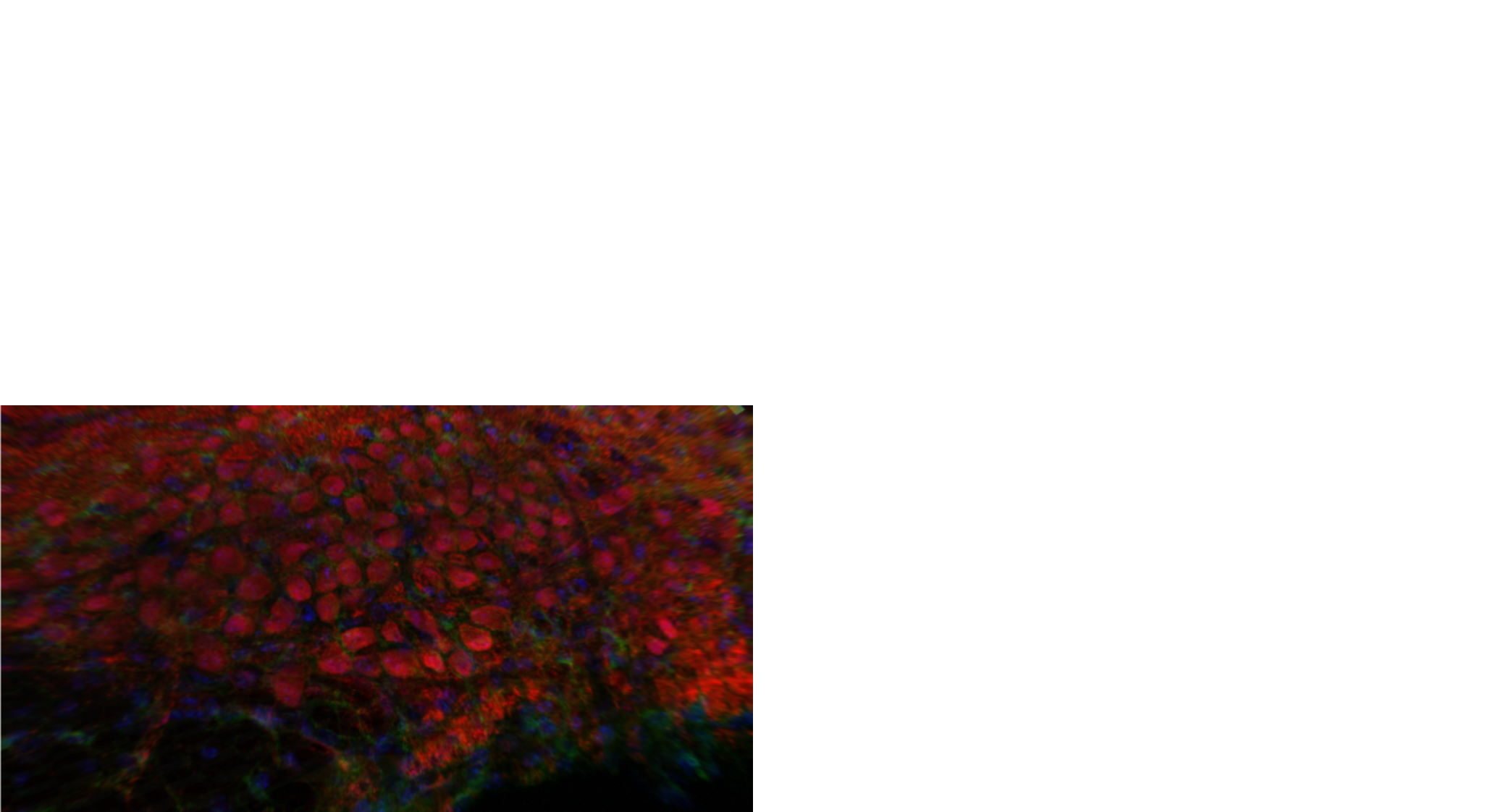}\label{simple_color}} & \subfloat[]{\includegraphics[width=1.5in]{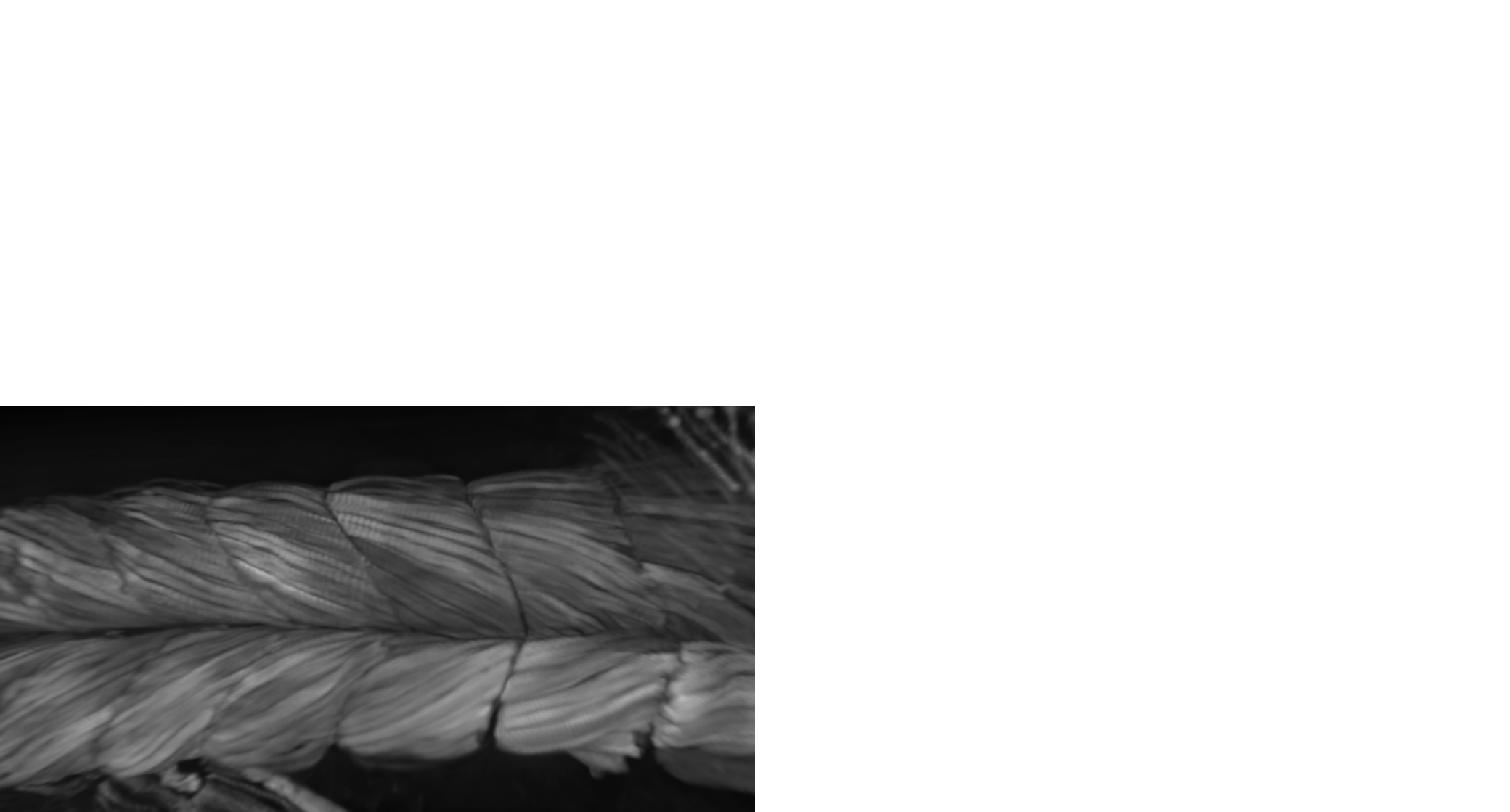}\label{simple}}  \\
\subfloat[]{\includegraphics[width=1.5in]{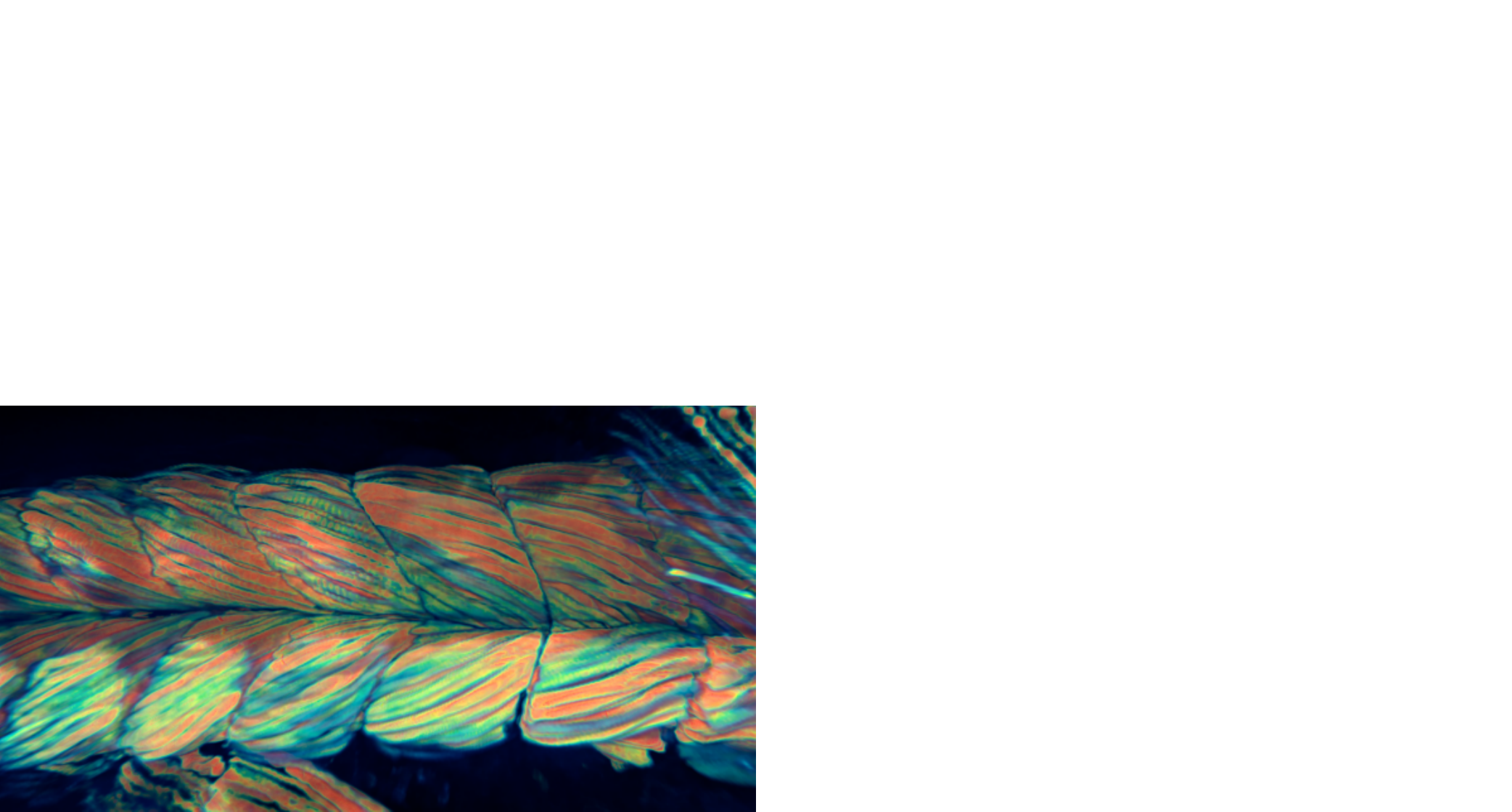}\label{color}} & \subfloat[]{\includegraphics[width=1.5in]{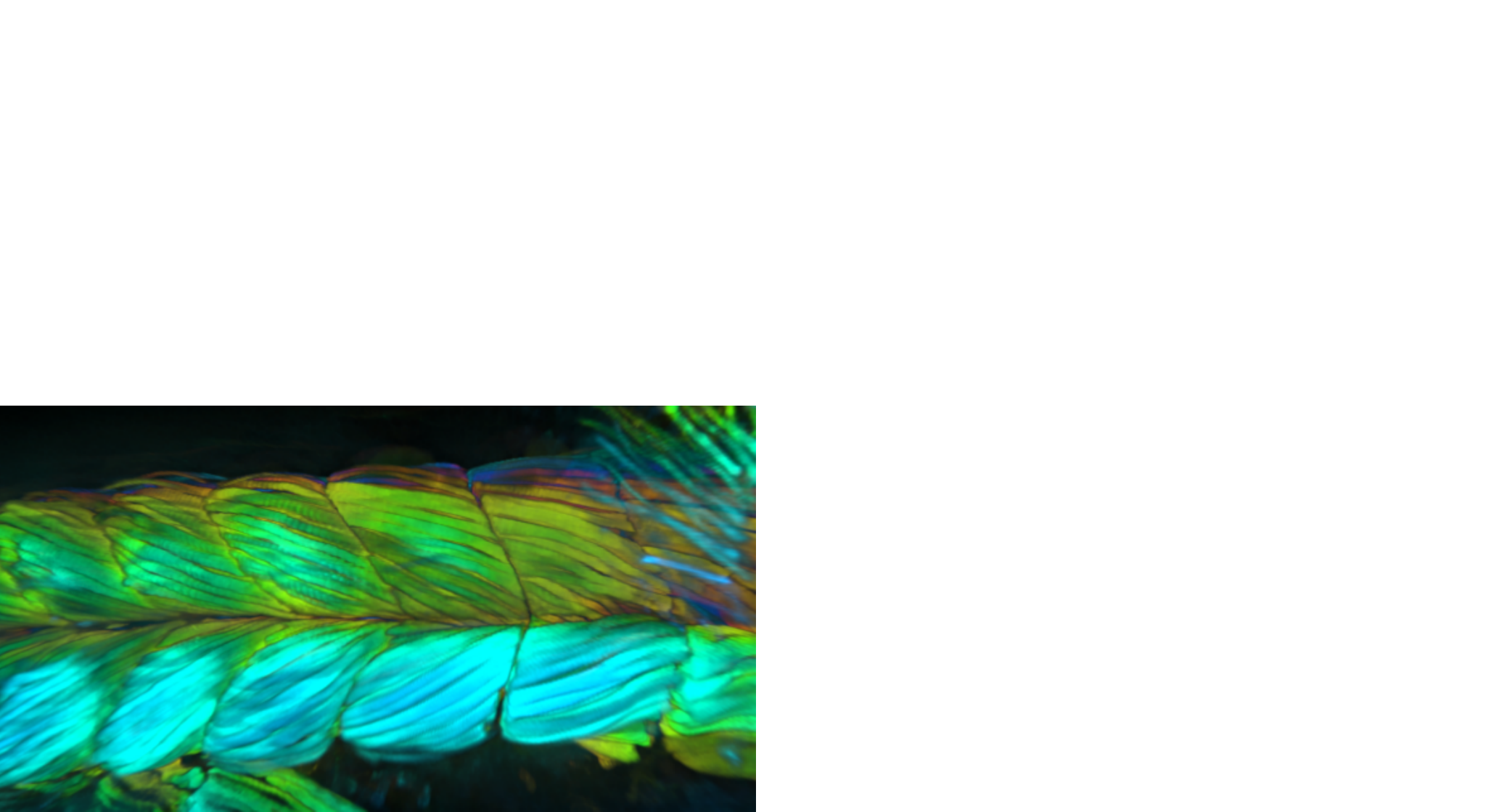}\label{depth}} \\
\end{tabular} 
\subfloat[]{\includegraphics[width=3.2in]{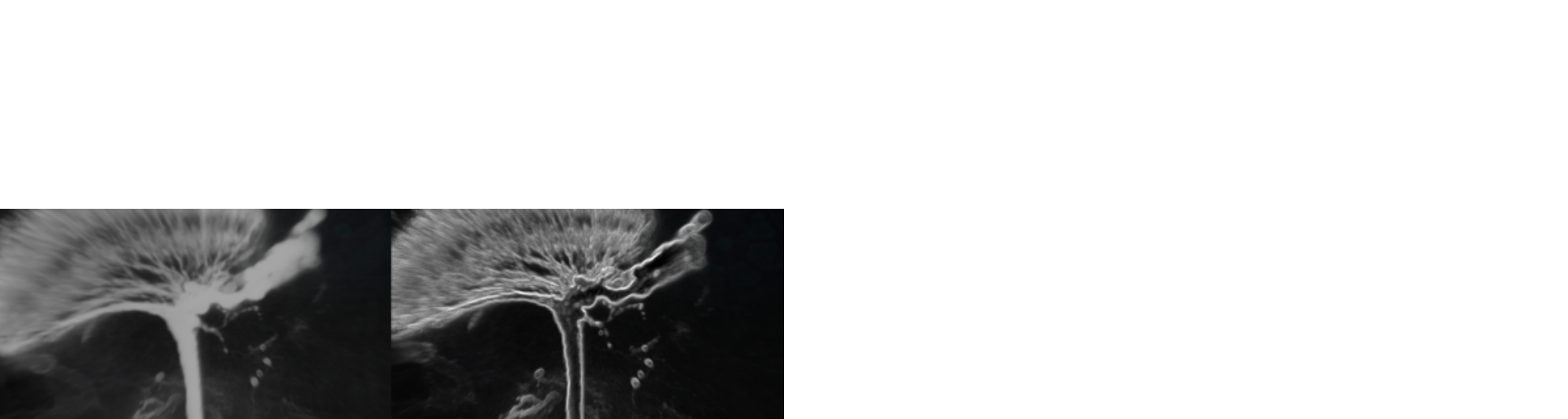}\label{grad}}
\caption{\textbf{Visualization enhancements.} Examples of visualization enhancements with different transfer functions showing: a) Multi-channel data; b) Linear scaling; c) Mapping volume data to a color palette; d) Mapping depth of the voxel in the volume to a color palette; e) Using gradients (of the left rendered volume) to emphasize boundaries in the right rendered volume. (b), (c), and (d) are data of zebrafish musculature kindly provided by Dr. A. Wood and Dr. P. Currie, Monash University.}
\label{fig:tf}
\end{figure}

Optical models for DVR require the intensity and opacity associated with every voxel. The hosted medium $f(x)$ represents some property of the tissue that has been imaged, e.g., the concentration of a marker, the response to a magnetic field, etc. DVR provides the flexibility to map data values to suitable optical properties depending on the visualization needs. This is done by appropriately defining the transfer functions $\tau[\cdot]$ and $T[\cdot]$, which map values of the medium to the intensity of emitted light and the opacity, respectively. The simplest case is just linear mapping, which can be useful for a broad range of data, see Figure~\ref{fig:tf}\subref{simple}. Intensity can be multi-channel, allowing for color images. Figure~\ref{fig:tf}\subref{simple_color} shows the rendering of a multichannel volume. If the source volume is single-channel, then it can be mapped to a color palette, as shown in Figure~\ref{fig:tf}\subref{color}. In some cases, the rendering may be more informative if the color represents additional information, such as depth of the voxel in the volume, 
like in Figure~\ref{fig:tf}\subref{depth}. Transfer functions can be arbitrary, so they can act on the gradient module $|\nabla f(x)|$, instead of $f(x)$, for instance. Taking into account the gradient of the scalar field of the medium may be useful to highlight its boundaries, like in Figure~\ref{fig:tf}\subref{grad}.

%Describe how different transfer functions allow to better visualize data.

%\begin{itemize}
%  \item{I would add this as it's own section as ``popped'' out from section 5}
%  \item{Use this to talk less about their technical grounding and more about how they're helpful in this situation}
%  \item{Talk also about color changing control which was some effort and is really useful for the scientists to use}
%  \item{Talk about how this control over the rendering method allows for the user to have fine detail of control on how the data is presented and why}
%  \item{I would put the pretty pictures here instead}\
   
%\end{itemize}

%%% Local Variables:
%%% mode: latex
%%% TeX-master: "syglasspaper-main"
%%% End:

\section{Conclusion}

Exploration and analysis of large-scale bioimages in virtual reality with HMD's
is challenging, but the benefits may exceed expectations with a fully developed technology. There is still a long way to go. The optimal set of tools to be developed in VR is still unknown. The best UI experience for using them has not yet been defined. There seem to be a high demand for analytical tools that can work semi-automatically, like volume auto-segmentation, but how to best operate these tools in immersive VR remains an open question.

%\begin{itemize}
%\item{Re hash the proposal section from the introduction breifly}
%\item{List our future additions we hope to make}
%\item{potentials could be:}
%  \begin{itemize}
%  \item{improve the range of input file types that we can ingest?}
%  \item{create automatic tools to complement the manual ones?}
%  \end{itemize}
%\end{itemize}

%%% Local Variables:
%%% mode: latex
%%% TeX-master: "syglasspaper-main"
%%% End:

\section*{Acknowledgment}
This research was supported in part by grants NIH / NIGMS-U54GM104942, NIH-1-R21-DC012638, NIH-5-R01-DC007695, and WVU Straton Research Chair Funds. We are also grateful for the contributions provided by Nathan Spencer, Jennifer Nguyen, Haofan Zheng, and Jordan Brack.

\bibliographystyle{IEEEtran}
\bibliography{syglasspaper-main}

\begin{IEEEbiographynophoto}{Stanislav Pidhorskyi}\\
    (\texttt{stpidhorskyi@mix.wvu.edu}) is with the Lane Department of Computer Science and Electrical Engineering of West Virginia University.
\end{IEEEbiographynophoto}
\begin{IEEEbiographynophoto}{Michael Morehead}\\
   (\texttt{mmorehea@mix.wvu.edu}) is with the Lane Department of Computer Science and Electrical Engineering of West Virginia University.
\end{IEEEbiographynophoto}
\begin{IEEEbiographynophoto}{Quinn~Jones}\\
   (\texttt{qjones1@mix.wvu.edu}) is with the Lane Department of Computer Science and Electrical Engineering of West Virginia University.
\end{IEEEbiographynophoto}
\begin{IEEEbiographynophoto}{George~Spirou}\\
   (\texttt{gspirou@hsc.wvu.edu}) is with the Rockefeller Neuroscience Institute of West Virginia University.
\end{IEEEbiographynophoto}
\begin{IEEEbiographynophoto}{Gianfranco~Doretto}\\
   (\texttt{gidoretto@mix.wvu.edu}) is with the Lane Department of Computer Science and Electrical Engineering of West Virginia University.
\end{IEEEbiographynophoto}

\end{document}